\pdfoutput=1
\RequirePackage{ifpdf}
\ifpdf
\documentclass[pdftex]{sigma}
\else
\documentclass{sigma}
\fi

\usepackage{cite}

\numberwithin{equation}{section}

\def\d{\delta}

\def\e{\epsilon}

\def\G{\Gamma}

\def\m{\mu}
\def\n{\nu}

\def\r{\rho}
\def\s{\sigma}

\def\t{\tau}

\def\a{\alpha}
\def\b{\beta}\def\th{\theta}

\def\p{\partial}

\begin{document}
\allowdisplaybreaks

\renewcommand{\thefootnote}{$\star$}

\renewcommand{\PaperNumber}{013}

\FirstPageHeading

\ShortArticleName{Semiclassical Loop Quantum Gravity and Black Hole Thermodynamics}

\ArticleName{Semiclassical Loop Quantum Gravity\\ and Black Hole Thermodynamics\footnote{This
paper is a~contribution to the Special Issue ``Loop Quantum Gravity and Cosmology''.
The full collection is available at
\href{http://www.emis.de/journals/SIGMA/LQGC.html}{http://www.emis.de/journals/SIGMA/LQGC.html}}}

\Author{Arundhati DASGUPTA}

\AuthorNameForHeading{A.~Dasgupta}

\Address{University of Lethbridge, 4401 University Drive, Lethbridge T1K 7R8, Canada}
\Email{\href{mailto:arundhati.dasgupta@uleth.ca}{arundhati.dasgupta@uleth.ca}}

\ArticleDates{Received March 22, 2012, in f\/inal form February 05, 2013; Published online February 16, 2013}

\Abstract{In this article we explore the origin of black hole thermodynamics using semiclassical
states in loop quantum gravity.
We re-examine the case of entropy
using a~density matrix for a~coherent state and describe correlations across the horizon due to
${\rm SU}(2)$ intertwiners.
We further show that Hawking radiation
is a~consequence of a~non-Hermitian term in the evolution operator, which is necessary for entropy
production or depletion at the horizon.
This non-unitary
evolution is also rooted in formulations of irreversible physics.}

\Keywords{black holes; loop quantum gravity; coherent states; entanglement entropy}

\Classification{83C57; 81Q35; 81T20}

\renewcommand{\thefootnote}{\arabic{footnote}}
\setcounter{footnote}{0}

\section{Introduction}

With new results in lattice gravity, loop quantum gravity (LQG) and various other approaches to
quantum gravity, there is hope that we might f\/ind the quantum of
space-time~\cite{qg,mbh,abck,lqg,lqg1,lqg2}.
Though unifying gravity with other forces of nature will remain a~puzzle.
General theory of relativity is a~theory of gravity and quantum f\/ield theory is the theory of
particle dynamics and both are experimentally verif\/ied.
Yet when we describe quantum f\/ields in gravitational backgrounds, new phenomena arise which seem to
signify the existence of new physics.
Black hole thermodynamics and Hawking radiation are particularly interesting phenomena.
The laws of black hole mechanics derived in~\cite{bch} were the f\/irst indication that the black
hole might be a~system similar to a~thermodynamic ensemble.
The area increasing theorem~\cite{hawk} led to the conjecture that the entropy of the thermodynamic
system is proportional to the area of the horizon~\cite{bh}.
The area of the horizon divided by four Planck length squared today is known as the
Bekenstein--Hawking entropy.
The use of quantum f\/ields near the horizon then showed that black holes emit radiation in thermal
spectrum at a~temperature given by the surface gravity of the black hole~\cite{hawk2}, and the
Planck's constant naturally got included in the quantif\/ication of entropy, temperature etc.
This conf\/irmed that there is a~`quantisation' of the space-time and the microscopic structure
will explain the origin of thermodynamic quantities.
There are several explanations of the nature of the microscopic structure of the black hole entropy
\cite{mbh,abck}, and the origin of thermodynamics, some of which will be reported in this volume.
In this article I~shall discuss the use of `coherent states' in LQG to derive the physics of the system
\cite{hall,thiemwinkl}.
The use of coherent states is well known in quantum electrodynamics in particular for lasers, and
we will use similar coherent states for gravity.
Once the coherent states are identif\/ied for the black hole, the derivation of thermodynamics
follows the standard entropy formulation of correlated systems, part of which has been traced away~\cite{adg2}.
A reduced density matrix is derived for a~region outside the horizon and the wavefunction inside
the horizon is traced over.
This reduced density matrix gives an entropy proportional to the area of the horizon.
There are semiclassical corrections to the Bekenstein--Hawking term, and the nature of the
corrections is a~function of the discretisation or graph embeddings used to describe the system~\cite{adg3}.
In this article we describe the derivation of entropy using coherent states,
and discuss brief\/ly the nature of correlations which arise due to the gauge invariant coherent
states using ${\rm SU}(2)$ intertwiners.
These calculations are a~preliminary indication of what a~`physical coherent state' might carry as
correlations at the horizon~\cite{adg4}.
The correlations make the state inside the horizon `entangled' with the state outside the horizon.

As the coherent state is def\/ined in the canonical formalism where space-time is foliated with
constant time slices, the coherent state in discussion is def\/ined in one time slice, which we take
as the initial time slice.
We then use a~semi-classical Hamiltonian to evolve the coherent state in time.
The derivation of time evolution is trivial if we restrict the evolution operator to be unitary.
Non-trivial evolution occurs if the classically forbidden regions behind the horizon get exposed
using a~non-Hermitian term added to the Hamiltonian which facilitates entropy production~\cite{adg4a,adg5}.
The change in entropy of the black hole is carried away as Hawking f\/lux which escapes
to inf\/inity.
This time evolution is irreversible in nature and is very similar to such time evolutions in
complex systems~\cite{prigogine}.

In the next section we brief\/ly describe coherent states for gravity.
We also review the concept of entanglement entropy in this section.
The third section describes the origin of entropy using a~density matrix.
It includes a~description of the tracing mechanism when the coherent state is def\/ined using
intertwiners at vertices.
In the fourth section we describe time evolution and the relation of this entropy production
process to physics of complex systems.
The f\/ifth section discusses the results and the open problems yet to be solved using this formalism.

\section{Coherent states in loop quantum gravity}

The coherent states are useful semiclassical states in a~quantum theory.
In case of the simple harmonic oscillator or in quantum electrodynamics, these can
be formulated as the eigenstate of the annihilation operator.
Explicitly
\begin{gather*}
\hat a|\psi\rangle=z|\psi\rangle,
\end{gather*}
where $\hat a$ is the annihilation operator for the positive energy modes of the theory and $z$ is
the eigenvalue.
These are usually Gaussians, e.g.\
for the harmonic oscillator
\begin{gather*}
\psi= \frac{1}{\sqrt{2\pi \hbar} }\exp\left(-\frac{(x-z)^2}{2 \hbar}\right),
\end{gather*}
where $z=x_0+i p_0$ is a~point in the classical complexif\/ied phase space, $\hbar=h/2\pi$,
and $h$~is Planck's constant.
Thus the coherent state is a~Gaussian `peaked' on the classical phase space point. In these coherent states the expectation values of operators are
obtained as their classical values and trajectories also evolve along classical paths, preserving
the coherence.

The coherent states appear as the kernel of a~transformation from the Hilbert space
$L^2({\mathbb R})$ to the Segal--Bergmann representation of the wave functions or $H({\mathbb C})\cap L^2({\mathbb C})$~\cite{hall}.
Using the def\/inition of the coherent state as a~kernel Hall~\cite{hall} generalised the
coherent states to obtain those def\/ined for a~${\rm SU}(2)$ Hilbert space.
These appear as kernels in the transformation from the ${\rm SU}(2)$ Hilbert space to the intersection of
the Hibert space def\/ined in the complexif\/ied ${\rm SU}(2)$ phase space which incidentally is ${\rm SL}(2,{\mathbb C})$ space.
These are therefore often named as `complexif\/ier coherent states'.
In a~particular def\/inition of the kernel, it is obtained as
\begin{gather}\label{eqn:com}
K(h,g)= e^{\frac t2\Delta}\delta(h',h)_{h'\rightarrow g},
\end{gather}
where $h$ is a~${\rm SU}(2)$ element, and $g$ is a~${\rm SL}(2,{\mathbb C})$ element,
$\Delta$~is the Laplacian on the group manifold,
$t$~is a~parameter.
The Hall coherent states can be directly applied to gravity as written in terms of LQG variables
gravity is a~${\rm SU}(2)$ theory~\cite{thiemwinkl}.

We begin by writing gravity in terms of LQG variables.
The phase space of loop quantum gravity is obtained using a~canonical formulation of gravity.
The manifold is taken dif\/feomorphic to $\Sigma\times {\mathbb R}$.
The $\Sigma$ slices
foliate the space-time, and using the ADM metric, the dynamical variables are identif\/ied as the
intrinsic metric $q_{ab}$ of the slice and the extrinsic curvature
$K_{ab}$ with which the slices are embedded in the space-time.
The intrinsic metric can be written in terms of the tangent space triads or soldering forms
$e^I_a$.
The phase space for LQG is again a~redef\/inition of these and described thus~\cite{lqg,lqg1,lqg2}
\begin{gather}
A_a^I =\G_a^I - \beta K_{ab} e^{I b}, \qquad E^a_I = \frac1{\beta} (\det  e) e^a_I,
\label{defn}
\end{gather}
where $e_a^I$ are the usual triads the `square root' of the metric $q_{ab}=e_a^I e_b^I$, $K_{ab}$ is
the extrinsic curvature, $\G_a^I$~are the associated spin connections to the triads, $\beta$ is the one
parameter ambiguity,
which is known as the Immirzi parameter.
The quantisation of the Poisson algebra of these variables is done by smearing the connection along
one dimensional edges $e$ of length~$\delta_e$ of a~graph $\G$ to get holonomies~$h_e(A)$.
The triads are smeared in a~set of 2-surface decomposition of the three dimensional spatial slice
to get the
corresponding momentum~$P_e^I$. (The momenta are also labeled by the edges as each edge intersects
the corresponding 2 surface once by construction,
or every edge has a~corresponding unique two surface and vice versa.)
The regularised LQG variables are
\begin{gather}\label{eqn:regul}
h_e(A) = {\cal P} \exp\left(\int_e A\right), \qquad P_{e}^I = \int_S * E^I.
\end{gather}

The algebra is then represented in a~kinematic `Hilbert space', in which the physical constraints
have been `formally' realised.
Once the phase space variables have been identif\/ied, one can write
a coherent state for these~\cite{hall}, i.e.\
minimum uncertainty states peaked at classical values of $h_e$, $P^I_e$ for one edge of the graph.
The tensor product of these coherent states for each edge of the graph gives the coherent state for
the entire space-time.
Using the Peter--Weyl theorem for the delta function $\delta(h,h')=\sum_j(2j+1)\chi_j
\big(h'h^{-1}\big)$, where $\chi_j$ is the character of the $j$th representation of ${\rm SU}(2)$, the complexif\/ier
coherent state~\eqref{eqn:com} can be written as
\begin{gather}
\tilde\psi^t(g_e)= \sum_j (2j+1) e^{-tj(j+1)/2} \chi_j \big(g_e h_e^{-1}\big).\label{eqn:coh}
\end{gather}
These are coherent states peaked at the classical holonomy with a~width controlled by the parameter~$t$.
A `momentum representation' of this is obtained by def\/ining a~Fourier transform which is such that
the `momentum eigenstate' $|jmn\rangle $ gives
\begin{gather*}
\langle h|jmn\rangle =\pi_j(h)_{mn}, \qquad \hat P_e|jmn\rangle = j(j+1) \hbar |jmn\rangle.
\end{gather*}
The $|jmn\rangle $ are thus the usual basis spin network states given by $\sqrt{2j+1}\pi_j(h_e)_{mn}$
(normalised in the $h_e$ basis), which is the $mn$ element of the
$j$th irreducible representation of the~${\rm SU}(2)$ mat\-rix~$h_e$, and $\hat P_e= \sqrt{\hat P_e^I \hat P_e^I}$.
Taking the scalar product of the conf\/iguration coherent state~\eqref{eqn:coh} with the momentum
eigenstate gives the momentum `coherent state' as $\psi^{t}_e=e^{-tj(j+1)/2}\pi_j(g_e)_{jmn}$, where
 $\pi_j(g_e)_{mn}$ is the $j$th irreducible representation of the ${\rm SL}(2,{\mathbb C})$ element~$g_e$.
Thus the cohe\-rent state in the momentum representation is
\begin{gather*}
|\psi^t(g_e)\rangle = \sum_{jmn} e^{-t j(j+1)/2} \pi_j (g_e)_{mn}|jmn\rangle.
\end{gather*}
In the above $g_e$ is a~complexif\/ied
classical phase space element $e^{i T^I P^{I\rm cl}_e/2}h^{\rm cl}_e$, where $P^{I\rm cl}_e$
and~$h^{\rm cl}_e$ represent classical momenta and
holonomy
obtained by embedding the edge in the classical metric and~$T^I$ are  ${\rm SU}(2)$ generators.
The $j$ is the
quantum number of the ${\rm SU}(2)$ Casimir operator in that representation, and $m$, $n$ represent azimuthal
quantum numbers which run from $-j,\dots,j$.
Similarly, $(2j+1)\times(2j+1)$-dimensional representations of the $2\times2$ mat\-rix~$g_e$ are
denoted as $\pi_j(g_e)_{mn}$.
The coherent state is precisely peaked with maximum probability at the $h^{\rm cl}_e$ for the
variable $h_e$ as well as the classical momentum~$P_e^{I\rm cl}$ for the variable~$P_e^{I}$.
The f\/luctuations about the classical value
are controlled by the parameter~$t$ (the semiclassicality parameter).
This parameter is given by $l_p^2/a$ where $l_p$ is Planck's constant and~$a$ is a~dimensional constant which characterises the system.
For the Schzwarzschild black hole $a$ can be taken as proportional to the area of the horizon.
The coherent state for an entire slice can be obtained by taking the tensor product of the coherent
state for each edge which form a~graph~$\G$,
\begin{gather}
\Psi_{\G}= \prod_e \psi^t_e.
\label{coh}
\end{gather}

{\it Thus we are considering a~semiclassical state, which is a~state such that expectation values
of operators are closest to their classical values.
The information of the classical phase space variables are encoded in the
complexified ${\rm SU}(2)$ elements labeled as~$g_e$.
The fluctuations over the classical values are controlled by the semiclassical
parameter~$t$.}

The density matrix which describes the entire black hole slice is obtained as
\begin{gather*}
\rho^{\rm total}= |\Psi_\G\rangle\langle \Psi_\G|,
\end{gather*}
where $|\Psi_\G\rangle $ is the coherent state wavefunction for the entire slice, a~tensor product of
coherent state for each edge.
These coherent states have
been studied in various other forms~\cite{livine,bianchi}.
In this paper we
also discuss the ${\rm SU}(2)$ invariant coherent state described in~\cite{thiemwinkl} using
intertwiners at the vertices.
We will describe the detailed derivation of the intertwiners
for a~particular graph in the next section using this generalisation.
We should mention that previously other semiclassical states had been considered in the framework
LQG
known as weave states~\cite{weave}, but none have been yet used to describe black hole entropy.
Coherent states have been used to describe quantum cosmology and
FLRW universes~\cite{frw}.
Relatively recent reviews on approaches to black hole entropy and corrections to black hole entropy
in LQG can be found \mbox{in~\cite{rev1,rev2,rev3}}.
A review of black hole entropy in LQG and recent description of two dimensional black hole
evaporation can be found in~\cite{rev4}.
Before we begin the discussion on derivation of entropy using density mat\-ri\-ces, we brief\/ly describe
the concept of `entanglement entropy' and its use in quantum f\/ield theory in curved space-time to
describe entropy of black holes.
Our derivation will be totally `gravitational in origin' def\/ined using the LQG phase space
variables.

\section{Entanglement entropy}
In quantum f\/ield theory, we use the def\/inition of entropy due to von Neumann.
It is def\/ined thus given a
density matrix $\rho$
\begin{gather*}
S= - \operatorname{Tr}(\rho\ln\rho).
\end{gather*}
If the state is pure the entropy is zero, if the state is mixed the entropy is obtained as non-zero.
This could be a~system in which part of the wavefunction or the state of the system has been traced away.
The tracing can be done for systems which have a~product Hilbert spaces: e.g.\
$H=H_I\otimes H_O$.
The $H_I$ is the Hilbert space which def\/ines the `internal space', $H_O$ is the outside Hilbert
space.
The basis states also thus appear in the product structure $|i\rangle \otimes|j\rangle $.
The states are said to be entangled if the coef\/f\/icients of the basis states do not factorise, e.g.
\begin{gather*}
|\psi\rangle= \sum_{ij} d_{ij}|i\rangle |j\rangle
\end{gather*}
is entangled if
\begin{gather}\label{eqn:entanglement}
d_{ij}\neq d_i d_j.
\end{gather}
A density matrix def\/ined thus
\begin{gather*}
\rho=|\psi\rangle\langle \psi|
\end{gather*}
can have a~partial trace performed on it,
\begin{gather*}
\operatorname{Tr}\rho= \sum d^*_{ij}d_{ji'}|i\rangle\langle i'|,
\end{gather*}
and the entropy computed using the above.
We will now describe the `entanglement entropy' of quantum f\/ields in a~black hole background.
This shed light on why the entropy was proportional to the area of the horizon and not the volume of
the black hole.
A review on entanglement entropy of black holes can be found in~\cite{solud}.
Srednicki computed the entanglement entropy of a~quantum f\/ield, where the f\/ield enclosed in a
sphere of radius `$R$' is traced out.
The entropy of the remaining space-time is proportional to the area of the sphere~\cite{sred}.
This was a~very interesting calculation however doing a~similar calculation for a~black hole
background did not give f\/inite answers.
The reason is that the `metric' in the coordinates of an outside observer is singular at the
horizon.
The near horizon volume is inf\/inite, and thus the entropy due to the quantum f\/ields in this volume
is also inf\/inite.
't~Hooft~\cite{bw} introduced a~mass dependent `cutof\/f' as if a~brickwall was steeling of\/f the
quantum f\/ields, which satisf\/ied a~Dirichlet boundary condition at the wall.
The presence of the brick wall could be used to obtain a~f\/inite value for the entropy.
We shall brief\/ly review the procedure here.

\subsection{The brickwall model}

The scalar f\/ield $\phi$ is taken in the background of a~Schwarzschild black hole whose metric in
spherical coordinates is
\begin{gather}\label{eqn:metric}
ds^2 = -\left(1-\frac{2M}{r}\right) dt^2 + \left(1-\frac{2M}{r}\right)^{-1} dr^2 + r^2\big(d\theta^2 +
\sin^2\theta d\phi^2\big),
\end{gather}
where $M$ is the mass of the black hole, $G=1$, $c=1$.

The scalar f\/ield's Klein--Gordon equation in this background is
\begin{gather*}
\frac{1}{\sqrt{-g}}\partial_{\mu}\left(\mathop{\sqrt{-g}} g^{\m \n}\partial_{\nu}\phi\right) -m^2=0.
\end{gather*}

Using the inverse of the metric~\eqref{eqn:metric} one f\/inds that the Klein--Gordon equation is
\begin{gather*}
\left(1-\frac{2M}{r}\right)^{-1} \left({-\partial_t^2}\right)\phi + \frac1{r^2} \partial_r
\left(r^2 \left(1-\frac{2M}{r}\right)\partial_r\right)\phi
+ \frac{1}{r^2}L^2\phi -m^2\phi=0,
\end{gather*}
where $L^2$ is the angular momentum operator.
The above can be solved in the eikonal approximation using
\begin{gather*}
\phi= e^{i E t} e^{i \int k dr},
\end{gather*}
where
\begin{gather*}
k^2= \frac{r^2}{r(r-2M)}\left(\left(1-\frac{2M}{r}\right)^{-1} E^2 -\frac1{r^2} l(l+1) -m^2\right).
\end{gather*}
As evident the wavenumber is quite big near the horizon, justifying the eikonal approximation (the
phase dominates).
Due to the Dirichlet boundary condition, $\phi=0$ at the wall, a~particular value of the radius
taken as $r=2M+h$ and an outer boundary $L$, the
wavefunction's wavenumber gets quantised as integer multiples of~$\pi$
\begin{gather*}
\pi n = \int_{2M+h}^L k dr.
\end{gather*}
The total number of such eigenmodes with energy $E$ is therefore the sum over angular momentum
(each with $2l+1$) degeneracy
\begin{gather*}
\int (2 l+1) dl   \pi n= g(E).
\end{gather*}

Given this one can compute
\begin{gather*}
e^{-\beta F}= \prod_{n,l,l_3} \frac{1}{1 - e^{-\tilde\beta E}},
\end{gather*}
where $F$ is the free energy, and the righthand side is the thermal distribution, at a~temperatu\-re~$1/\tilde\beta$,
such that
\begin{gather*}
\pi \tilde\beta F =\int dg(E) \log \big(1-e^{\tilde\beta E}\big)=-\int_0^{\infty} dE
\frac{g(E)}{e^{\tilde\beta E}-1}.
\end{gather*}
Using the value of $g(E)$
\begin{gather*}
F= -\frac{1}{\pi}\int_0^\infty~dE~\int_{2M+h}^L dr \left(1-\frac{2M}{r}\right)^{-1} \int
\frac{(2l+1)}{\big(e^{\tilde\beta E} -1\big)} dl \sqrt {E^2- \left(1-\frac{2M}{r}\right)\left(
\frac{l(l+1)}{r^2}\right)}
\end{gather*}
(we use the approximation that the scalar f\/ields are massless).
$F$ is then approximated in~\cite{bw} as
\begin{gather*}
F= -\frac{2\pi^3}{45 h}\left(\frac{2M}{\tilde\beta}\right)^4 - \frac{2}{9\pi} L^3\int_0^\infty
\frac{dE E^3}{e^{\tilde\beta E}-1}.
\end{gather*}
Using the def\/inition of entropy
\begin{gather*}
S=\tilde\beta (U-F), \qquad U =\frac{\partial}{\partial\tilde \beta}\big(\tilde\beta F\big)
\end{gather*}
one obtains
$
S= 4 M^2$.
This is indeed proportional to the area of the horizon, but one had to put the cutof\/f radius
$h=\frac{1}{720\pi M}$, a~very particular value to achieve the result.

There are some observations about this derivation of entropy:
$(i)$~There is no entanglement of modes across the horizon of the QFT modes, what this represents is
the entropy of a~gas of scalar particles outside the horizon.
$(ii)$~Even though the cutof\/f is mass dependent the proper-distance of the brickwall from the horizon
is a~constant.
The question still remains why it is this constant and not another one.

\subsection{Scalar entanglement for a~bifurcate horizon}

It is very interesting that this `entropy of a~gas of scalar particles' using the brick wall method could be placed in an
`entanglement entropy' context.
In this formalism one traces over the `outside modes' to obtain a~density matrix and the entropy of
the resultant partially traced sector of the system.
The details of this derivation can be found in~\cite{barfr} and references therein.
The crux of the computation is hinged on the def\/inition of the wavefunction of the black hole as a
`path-integral' in a~space-time with a~complex metric.
The space-time is the complexif\/ied Kruskal extension of the Schwarzschild black hole.
The metric in spherical coordinates for the Schwarzschild black hole is
given in~\eqref{eqn:metric}.
One then def\/ines the Kruskal coordinates as
\begin{gather*}
U= - e^{-(t-r-2M\ln(r/2M-1))/4M},\qquad
V = e^{(t+ r + 2M\ln(r/2M -1))/4M},\\
UV = \left(1-\frac{r}{2M}\right)\exp\left(\frac{r}{2M}\right),
\end{gather*}
such that the Kruskal metric is
\begin{gather*}
ds^2 = \frac{-32 M^3}{r}\exp^{-\left(r/2M\right)} dU dV + r^2 d\Omega^2.
\end{gather*}
As evident the metric is extendible past $r=2M$, the horizon, and the horizon is bifurcate, with
$V=0$ coinciding with the past horizon and $U=0$ coinciding with the future horizon.
There are two asymptotics, and the transformations $U\rightarrow -U$ and
$ V\rightarrow -V$ map the two asymptotics.
One could then def\/ine the Lorentzian metric as the real section of a~complex metric, by def\/ining
the coordinates $U$ and $V$ w.r.t.\ a~`complex time' $z=\tau+i t$.
The $\tau$ is the Euclidean time, and it is such that
in the Euclidean section $-2\pi M<\tau<+2\pi M$.
The two asymptotics are then connected in some sense through this `Euclidean' section.
The complexif\/ied coordinates are
\begin{gather*}
U= -e^{(iz -i 2\pi M + r+2M\ln(r/2M-1))/4M},\qquad
V = e^{(-iz +i 2\pi M+ r + 2M\ln(r/2M -1))/4M}.
\end{gather*}
Clearly as the imaginary time $\tau$ changes from $-2\pi M$ to $2\pi M$, one travels from one
asymptotics to another given by $z^+=2\pi M+it$ and $z^-=-2\pi M+it$ ($-\infty<t<\infty$) in
complex time.

One def\/ines a~path-integral through the Euclidean throat which could be labeled as the propagator
connecting the two asymptotics,
with $\phi\equiv\phi_+$ in one and $\phi\equiv\phi_-$ in the other.

Def\/ine
\begin{gather}\label{eqn:heat}
\Psi(\phi_{-},\phi_{+})= \langle \phi_{-}|\exp(-\tilde\beta \hat H/2)|\phi_{+}\rangle.
\end{gather}
Thus the path-integral is over the Euclidean `throat' connecting the two Lorentzian asymptotics.
Given $\tilde\beta=8\pi M$ and $\hat H$ is a~Hamiltonian which in the linearised approximation
gives the usual Hamiltonian of matter f\/ields including scalar f\/ields.
The computation in~\cite{barfr} is given for scalar Hamiltonian using heat kernel techniques.
A density matrix is def\/ined thus
\begin{gather*}
\rho= \int {\cal D}\phi_{+} \Psi^*(\phi'_{-},\phi_{+})\Psi(\phi_{-},\phi_{+}).
\end{gather*}
Plugging $\Psi$ from~\eqref{eqn:heat} and using $\int{\cal D\phi_{+}}|\phi_{+}\rangle\langle \phi_{+}|=1$
this reduces to
\begin{gather}\label{eqn:scl}
\rho=\exp(\Gamma)\langle \phi_{-}|\exp(-8\pi M \hat H)|\phi_{-}\rangle,
\end{gather}
where the prefactor is introduced to preserve $\int{\cal D\phi_{-}}\rho=1$ or the trace condition.
The entropy of this density matrix is then computed using $S=-\operatorname{Tr}(\rho\ln\rho)$.
As the f\/ield is in one of the asymptotic regions, the Hamiltonian is obtained in the one-loop
approximation as the due to the scalar f\/ields solved in the classical black hole background.
Due to the def\/inition of the density matrix as the expectation value of the Hamiltonian at a
temperature $1/\beta$~\eqref{eqn:scl}, this calculation can be eventually mapped to the computation
of entropy of a~gas of the scalar particles near the horizon and it is
obtained as
\begin{gather}\label{eqn:ent}
S= \frac{4M^2}{45}\int_{2M}^{r_0}\frac{ r^3 dr}{(r-2M)^2}.
\end{gather}

Though this is proportional to the horizon area some aspects remain as the brickwall model.
The integral in~\eqref{eqn:ent} is divergent and thus the entropy is inf\/inite.
The exact nature of the entanglement is not clear however in the way entangled states are def\/ined
in~\eqref{eqn:entanglement}.
`Entanglement' entropy of scalar f\/ields in f\/lat space can be found where the entanglement is
obvious as in~\eqref{eqn:entanglement} in the following references~\cite{sred, sor}.

In the next section we will compute the `entanglement' entropy of a~black hole using
`non-perturbative' coherent states in a~`quantum gravity' formalism.
The variables are `regularised'
gravitational degrees of freedom.
The entropy computed is {\it purely gravitational in origin}.
The nature of entanglement is also specif\/ied clearly exactly similar to the discussion of
quantum mechanical entanglement in equation~\eqref{eqn:entanglement}.
The `internal' Hilbert space is traced out and the entropy is computed using the density matrix
def\/ined in the `outside' Hilbert space.
{\it The answer is finite and no cutoff is introduced.}

\section{Gravitational entropy}
In the def\/inition of the tensor product form of the coherent state~\eqref{coh}, the total coherent
state is a~tensor product of coherent states at each edge.
There is a~${\rm SU}(2)$ Hilbert space at each edge of the graph~$\Gamma$.
The total Hilbert space is thus $H=\otimes H_{e}$.
The question therefore naturally arises if the coherent state def\/ined in the basis of these Hilbert
spaces are entangled or simply independent? The structure, $\Psi=\prod_e\psi_e$ is such that the
states appears independent in the tensor product Hilbert spaces, but can be factorised as
per equation~\eqref{eqn:entanglement}.
However, the classical
geometry is interwoven from edge to edge due to Einstein's equations.
Thus even though not written in a~manifest way; for two adjacent edges~$e_1$ and~$e_2$,
$\psi_{e_1}$ is related to $\psi_{e_2}$ due to the classical solution at which these are
constructed to be peaked.
{\it We label
this entanglement due to classical geometry as classical correlations.
We then compute the entanglement of edges outside the horizon with those inside the horizon.}
The coherent states as introduced in~\eqref{coh} are `gauge covariant'.
The gauge ${\rm SU}(2)$ transformations act at the vertices and they are discussed in details in Appendix~\ref{appendixA}.
Introducing intertwiners at the vertices makes the coherent states `gauge invariant'.
They no longer transform due to the ${\rm SU}(2)$ transformations as the intertwiners map the vertices to
the trivial representation of ${\rm SU}(2)$.
These introduce further `entanglement' of the coherent states of edges ending/beginning at the same vertex.
These correlations arise as the type of spins assigned to the edges meeting at a~vertex are
restricted to ensure gauge invariance.
{\it These correlations introduced due to `intertwiners' are labeled as `quantum correlations'}.
We thus identify two types of
correlations.
$(i)$~The classical correlations which are due to the relation of the $g_e$ from one
edge to the next already discussed in~\cite{adg2}.
$(ii)$ Quantum correlations due to the intertwiners which link the quantum spins of the edges meeting at a~vertex.
We discuss these correlations for the f\/irst time in this article.

\subsection{Classical correlations}
The classical correlations can be identif\/ied easily in a~particular slicing of the classical
geometry.
We discuss a~particular time slicing of the metric
where the intrinsic curvature of the time slices is f\/lat.
One such metric which has the time slices as f\/lat is the Lemaitre metric
\begin{gather*}
ds^2 = -d\tau^2 + \frac{dR^2}{\big[\frac{3}{2r_g}(R-\t)\big]^{2/3}} +
\left[\frac{3}{2}(R-\t)\right]^{4/3} r_g^{2/3}\big(d\theta^2 +\sin^2\theta d\phi^2\big).
\end{gather*}
The $r_g=2GM$  (in units of $c=1$) and
in the $\tau= {\rm const}$ slices one can def\/ine the induced metric in terms of a~`$r$' coordinate
def\/ined as $dr=dR/\left[3/2r_g(R-\tau_c)\right]^{1/3}$ ($\tau=\tau_c$ is a~constant) on the slice.
One gets the metric of the three slice to be
\begin{gather}\label{sph}
ds_3^2= dr^2 + r^2 \big(d\theta^2 + \sin^2\theta d\phi^2\big).
\end{gather}
The entire curvature of the space-time metric is contained in the extrinsic curvature or
$K_{\m\n}=\frac12\partial_\tau g_{\m\n}$ tensor of the $\tau= {\rm const}$ slices.
Now if there exists an apparent horizon somewhere in the above spatial slice, then that is located
as
a solution to the equation
\begin{gather*}
\nabla_a S^a + K_{ab}S^aS^b -K=0,
\end{gather*}
where $S^a$ ($(a,b=1,2,3)$ denote the spatial indices) is the normal to the horizon, $K_{ab}$ is the
extrinsic curvature in the induced
coordinates of the slice, and $K$ is the trace of the extrinsic
curvature.
If the horizon is chosen to be the 2-sphere, then in the coordinates
of~\eqref{sph}, $S^a\equiv(1,0,0)$,
the apparent horizon equation as a~function of the metric reduces to
\begin{gather}\label{classical}
K_{rr}\big(1-q^{rr}\big) -K_{\phi\phi}q^{\phi \phi} - \G^{\phi}_{\phi r} -K_{\th\th}q^{\th
\th}-\G^{\th}_{\th r}=0.
\end{gather}
In this article
we provide a~very simplif\/ied explanation of why this equation provides correlations across the
horizon.
Note that the f\/irst term of the equation disappears trivially as $1=q^{rr}$ for any point in the
spatial slice and not just at the
horizon in the classical metric.

If we plug in the exact values of the above in terms of Schwarzschild variables, we f\/ind that the
terms cancel groupwise at the horizon due to spherical
symmetry.
Thus the apparent horizon equation which respects spherical symmetry can be imposed simply as the
equation in the $\theta$ variable or the
$\phi$ variable
\begin{gather}\label{eqn:trunc}
K_{\phi\phi}q^{\phi \phi} - \G^{\phi}_{\phi r} =\frac1r\left(1-\sqrt{\frac {r_g}{r}}\right)= 0, \qquad
K_{\th\th}q^{\th \th} - \G^{\th}_{\th r}=\frac1r\left(1-\sqrt{\frac {r_g}{r}}\right)=0.
\end{gather}

This `truncated' versions of the apparent horizon equation can be implemented on the coherent
states in the
classical limit.
Due to spherical symmetry, the other half of the horizon equation gets automatically satisf\/ied.
Note we are not using two dif\/ferent equations to replace the one equation in~\eqref{classical}, but
observing that due to spherical symmetry, ensuring that the equation is satisf\/ied for the $\theta$
indexed variables ensures that the other half of the equation will also be zero.
In other words the equations are proportional to each other.
To regularise the equation we observe
\begin{gather*}
\G^{\th}_{\th r}= \frac12 q^{\th \th}(q_{\th \th, r})= -\frac{1}{2} q_{\th \th} \big(q^{\th
\th}_{,r}\big)= \frac12 q_{\th \th} \frac1{\delta_{e_r}}\big(q^{\th \th}(v_1)-q^{\th \th}(v_2)\big).
\end{gather*}
In the above we wrote the Christof\/fel connection as a~dif\/ference equation, $v_1$ and $v_2$ are the
vertices of the graph and
$\delta_{e_r}$ is the length of the radial edge connecting the two vertices.

The above then translates to a~equation for the truncated apparent horizon equation
\eqref{eqn:trunc} as
\begin{gather}\label{eqn:appar}
q^{\th \th}(v_{\rm out})= q^{\th \th} (v_{\rm in})\big[2\delta_{e_r}K^I_{\th} e^{\th}_I (v_{\rm in}) +1\big],
\end{gather}
where the $v_{\rm out}$ vertices are outside the horizon and the $v_{\rm in}$ are vertices within
the horizon and $K_{\th\th}q^{\th\th}=K_{\th}^Ie^{\th}_I$ ($K_a^I=K_{ab}e^{bI}$).
We then embed a~graph with edges along the coordinate lines~$r$,~$\th$,~$\phi$ of the spherical
coordinate grid, and compute the regularised variables
of~\eqref{eqn:regul}.
The details of this can be found in~\cite{adg2}.
In terms of the regularised variables~\eqref{eqn:appar} appears as
\begin{gather}\label{eqn:diff1}
V P_{e_\th}^2 (v_{\rm out})= P_{e_\th}^2(v_{\rm in})\left[2\operatorname{Tr}\left( h_{e_{\th}}^{-1}(v_{\rm
in})T^I\beta\frac{\partial h_{e_\th}(v_{\rm in})}{\partial \beta} \right) P_{e_\th I}(v_{\rm in})+V
\right],
\end{gather}
where V is the volume operator and $P_{e}^2=P_e^I P_e^I$.
The volume operator appears due to the fact that it is the densitised operator $\sqrt q e^{a}_I$
which is regularised and $\sqrt q$ is regularised into the volume operator.
The volume operator in spherically
symmetric coordinates is written as $P_{e_H}\delta_{e_H}$ and thus it can be computed for this
system.
We also use a~regularisation introduced in~\cite{adg2} for the~$K_{a}^I$ operator
which uses a~`derivative' w.r.t.\ the Immirzi parameter.
Using the def\/inition in~\eqref{defn}, and the def\/inition of holonomy one gets $\partial
h_e/\partial\beta=\big(\int K^I_a T^I dx^a\big)h_e$.
And thus multiplying with~$h_e^{-1}T^I$ and taking trace identif\/ies the appropriate component of
the extrinsic curvature in the continuum limit.
This regularisation works for the purposes of the computations of expectation values as verif\/ied in~\cite{adg5}.
Note that constants which appear due to the specif\/ication of the graph edges (like $\delta_{e_r}$)
are suppressed in the above.
And thus it is suf\/f\/icient to take the following form
of the apparent horizon equation
\begin{gather}\label{eqn:diff}
V P_{e_\th}^2 (v_{\rm out})= P_{e_\th}^2(v_{\rm in})\left[2\operatorname{Tr}\left(h_{e_{\th}}^{-1}(v_{\rm
in})T^I \beta\frac{\partial h_{e_\th}(v_{\rm in})}{\partial \beta}\right) P_{e_\th I}(v_{\rm in})+V
\right].
\end{gather}

We use the apparent horizon equation to solve
for $P_{e_\th}(v_{\rm out})$ in terms of $P_{e_\th}(v_{\rm in})$.
The apparent horizon equation is further supplemented with a~restriction that the area induced on
the horizon using the radial edges sums up to the total
area of the horizon.
Which translated into the phase space operators means
\begin{gather}\label{eqn:area}
\sum_{e_H} P_{e_H} =A_H.
\end{gather}

\begin{figure}[t]
\includegraphics{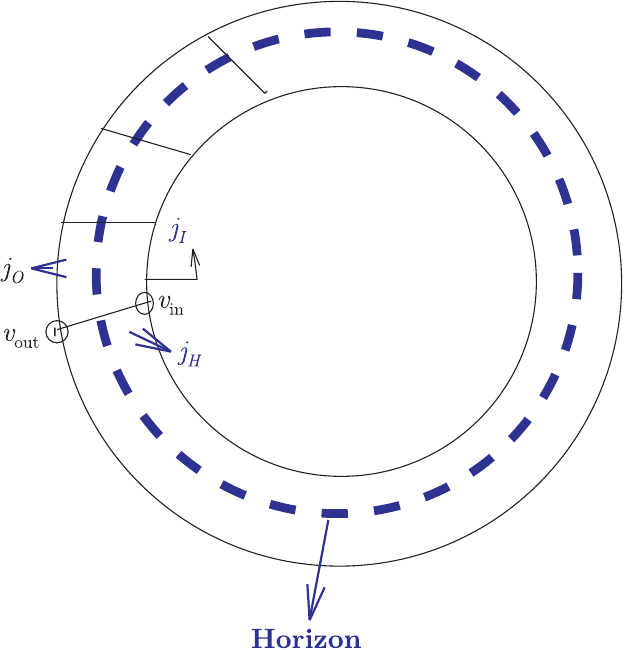}
\centering
\caption{Graph at the horizon.}
\label{fig:horizon}
\end{figure}

These~\eqref{eqn:diff}, \eqref{eqn:area} therefore specify the horizon information.
We then use these to perform the tracing mechanism on the density matrix and compute entropy.
Instead of computing the entropy with the entire density matrix, we concentrate on a~local set of
vertices surrounding the horizon.
As we saw, the apparent horizon correlates `angular edges'
immediately inside and outside the horizon.
We label this element $\rho^{\rm local}$ and isolate
it from the entire density matrix.
\begin{gather*}
\rho^{\rm total}= \rho^{\rm outside}\rho^{\rm local}\rho^{\rm inside},
\end{gather*}
where $\rho^{\rm local}$ covers a~band of vertices surrounding the horizon one set on a~sphere at
radius $r_g-\delta_{e_r}/2$ and one set
on a~sphere at radius $r_g+\delta_{e_r}/2$ within the horizon, as described in~\cite{adg2}, and
in the Fig.~\ref{fig:horizon} enclosed.
This local density matrix and the correlations due to the apparent horizon equation
\eqref{eqn:diff} was used
to derive entropy~\cite{adg2}.

We then further concentrate on two vertices $v_{\rm out}$ outside the horizon, $v_{\rm in}$ inside
the horizon which share a~radial edge $e_H$ and write a~density matrix for that $\rho^{\rm local}$.
At each vertex $v_{\rm out,in}$ there are two angular edges corresponding to the~$\theta$, $ \phi$
coordinate lines which are in\-going and two
angular edges which are outgoing.
These angular edges, 4 in number at each vertex have their classical~$g_{e}$ correlated with those
meeting at the other vertex.
Labeling $\psi_j(g)_{mn}=\frac{1}{\langle \psi|\psi\rangle^{1/2}}e^{-t/2j(j+1)}\pi_j(g_e)_{m n}$,
\begin{gather*}
\rho^{\rm local} = \sum_{\{j_{Oi}\}\{j_H \}\{j_{Ii}\},\{j'_{Oi}\}\{ j'_H\}
\{j'_{Ii}\}}\prod_{i=1}^{4}\psi_{j_{Oi}}(g_e(v_{\rm out})[g_e(v_{\rm
in})])_{m_{Oi}n_{Oi}}\psi_{j_H}(g_{e_H})_{m_H n_H} \\
\phantom{\rho^{\rm local}=}
\times
\prod_{i=1}^4\psi_{j_{Ii}}(g_e(v_{\rm in}))_{m_{Ii}n_{Ii}} \prod_{i=1}^4|\{j_{Oi}\}\{ j_{H}\}
\{j_{Ii}\}\rangle\langle \{j'_{Oi}\} \{j'_{H}\} \{j'_{Ii}\}|\\
\phantom{\rho^{\rm local}=}
\times
\prod_{i=1}^{4}\bar\psi_{j'_{Oi}}(g_e(v_{\rm out})[g_e(v_{\rm in})])_{m'_{Oi}n'_{Oi}}
\psi_{j'_H}(g_{e_H})_{m'_H n'_H}\prod_{i=1}^4\bar\psi_{j'_{Ii}}(g_e(v_{\rm in}))_{m'_{Ii}n'_{Ii}}.
\end{gather*}
$j_{Oi}m_{0i}n_{0i}$ label the spins of the edges at the outside vertex.
$j_{Ii}m_{Ii}n_{Ii}$ label the spins at the inner vertex and the $j_H$, $m_H$, $n_H$ label the spins
on the
radial edge connecting the two vertices.
We us~$\{j\}$ to label all the three indices~$jmn$.
The tracing over the modes inside the horizon gives in the $t\rightarrow0$ limit
\begin{gather*}
\rho^{\rm local}_{\rm reduced} = \sum_{\{j_{Oi}\}\{j_H \}\{j_{Ii}\} \{j'_{Oi}\}
\{j_H'\}}\prod_{i=1}^4\psi_{j_{Oi}}(g_e(v_{\rm out})(g_e(v_{\rm in})))_{m_{Oi}
n_{Oi}}\psi_{j_H}(g_{e_H})_{m_H n_H}\\
\phantom{\rho^{\rm local}_{\rm reduced}=}
\times
\prod_{i=1}^4\delta(P_{e_i}(v_{\rm in}),j_{Ii}t)
|\{j_{Oi}\} \{j_{H}\}\rangle\langle \{j'_{Oi}\} \{j'_{H} \}|\bar\psi_{j'_H}(g_{e_H})_{m'_H
n'_H}\\
\phantom{\rho^{\rm local}_{\rm reduced}=}
\times
\prod_{j'_{i}=1}^{4}\bar\psi_{j'_{Oi}}(g_e(v_{\rm out})[g_e(v_{\rm in})])_{m'_{Oi}n'_{Oi}}.
\end{gather*}

In the $t\rightarrow0$ limit, the diagonal terms dominate.
The density matrix collapses
to a~set of non-zero elements~\cite{thiemwinkl}.
This can be directly seen from
\begin{gather}
\frac{1}{\langle \psi|\psi\rangle}e^{-tj (j+1)}\pi_j(\bar g)_{mn}\pi_j(g)_{mn}\nonumber\\
\qquad
\approx
\frac{t^{3/2}}{4\sqrt{\pi}p} e^{-j p|m/j - (^Rp)_3/p|} e^{-j p |n/j -(^L p)_3/p|}
e^{-[(j+1/2)-p]^2/t}.\label{eqn:limit}
\end{gather}
In the $t\rightarrow0$ limit, this gets to be peaked like a~delta function at the values
$P=(j+1/2)t$, $m t,nt= {}^L P_3, {}^R P_3$ are
the left and right invariant momentum operators.
As in~\cite{adg2} the $j_{Oi}$ gets f\/ixed to be a~certain number due to the apparent horizon
equation~\eqref{eqn:diff}, and the
correspon\-ding~$m_{Oi}$, $n_{Oi}$ were also taken as f\/ixed numbers using a~`gauge f\/ixed' version of the
apparent horizon equation.
However this is slightly arbitrary as the classical data is gauge invariant.
We expect that when all the constraints including the
Gauss constraints are solved this will emerge as obvious.

The individual components of ${}^{L(R)}\hat P^3_{e_{Oi}}$, f\/ixes the $m_{0i}n_{0i}$ to
unique numbers.
Thus the outer classical data $j^{\rm cl}_{Oi}$, $m^{\rm cl}_{Oi}$, $n^{\rm cl}_{Oi}$
have to be specif\/ied completely to specify the `outer state'.
{\it In other words we can break the degeneracy of the density matrix with different~$m_{Oi}
n_{Oi}$ by making the
operator measurements.}

Now we come to a~subtle point I had not discussed before, and that is the fact that the part of the
radial edge $e_H$ (see Fig.~\ref{fig:horizon}) is hidden behind the horizon.
It thus makes sense to trace over that.
To implement that, we therefore divide the horizon edge~$e_H$ into two further edges~$e_{H1}$ and~$e_{H2}$.
We build the basis state for this using the fact that
\begin{gather*}
|jmn\rangle  = \sqrt{2j+1}\pi_{j}(h_e)_{mn}|0\rangle\\
\hphantom{|jmn\rangle}{}
=
\sum_{k}\sqrt{2j+1}\pi_j(h_{e_1})_{mk}\pi_j(h_{e_2})_{kn}|0\rangle
=\sum_k\frac{1}{\sqrt{2j+1}}|jmk\rangle |jkn\rangle,
\end{gather*}
where $|0\rangle$ is the vacuum state.
Splitting the horizon edge, and tracing over the internal half gives
\begin{gather}\label{horizontrace}
\operatorname{Tr}{e_2} |jmn\rangle\langle j'm'n'|
=\sum_{k'kn}\frac{1}{2j+1}|j'm'k\rangle\langle j'k'n'|jkn\rangle\langle j'm'k'|\nonumber\\
\phantom{\operatorname{Tr}{e_2} |jmn\rangle\langle j'm'n'|}
=\sum_{k}\frac{1}{2j+1}|jm k\rangle\langle jm'k|\delta_{nn'}.
\end{gather}

Similarly, for the irreducible representation components of the `wavefunction'
\begin{gather*}
\pi_j(g_e)_{mn}\bar \pi_j(g_e)_{m'n},
\end{gather*}
which multiplies the basis states in the density matrix, one can perform such a~splitting of the holonomy as
\begin{gather*}
g_e= e^{i T^I P^I} h_e= e^{i T^I P^I}h_{e_1} h_{e_2}=g_{e_1}h_{e_2},
\end{gather*}
and thus
\begin{gather*}
\sum_{kk'}\pi_j(g_{e_1})_{m k}\pi_j(h_{e_2})_{kn}\bar\pi_j(g_{e_1})_{m'k'}\bar\pi_j(h_{e_2})_{k'n}.
\end{gather*}

The sum over $n$ then gives
\begin{gather}\label{irrtrace}
\pi_j(g_{e_1})_{mk}\bar\pi_j(g_e)_{m'k}= \pi\big(e^{\iota T^I P^I}\big)_{mm'}.
\end{gather}
This and then its complex conjugate gives the delta function in the $t\rightarrow0$ limit peaked
at~$P_H$.
The
contribution from the horizon edge
to the density matrix is
\begin{gather*}
\frac{1}{2j_H+1}\sum_{m_H k_H}\delta(j_H t, P_{e_H})|j_H m_H k_H\rangle\langle j_H m_H k_H|.
\end{gather*}
This explains the origin of the $1/(2j_H+1)$ factor anticipated in~\cite{adg2}.

Thus the reduced density matrix `describing the vertex outside the horizon' is
\begin{gather*}
\mathop{\rm Lim}\limits_{t\rightarrow 0}\rho^{\rm local}_{{\rm reduced}}= \sum_{ m_H n_H} \frac{1}{2j_H+1}
\\
\hphantom{\mathop{\rm Lim}\limits_{t\rightarrow 0}\rho^{\rm local}_{{\rm reduced}}=}
\times
\prod_{i=1}^4|P_{e_{Oi}},
 {}^L p_{3 Oi} \,{}^R p_{3 Oi}\rangle |P_{e_H} m_H n_H\rangle\langle P_{e_H} m_H n_H|\langle P_{e_{Oi}},
 {}^L p_{3 Oi} \,{}^R p_{3 Oi}|.
\end{gather*}

And the tensor product density matrix becomes
\begin{gather*}
\rho^{\rm local}_{\rm reduced}= \prod_{v_{\rm out}}\sum_{m_H n_H}\frac{1}{2j_H+1}
\\
\phantom{\rho^{\rm local}_{\rm reduced}=}
\times
\prod_{i=1}^4|P_{e_{Oi}}, {}^L p_{3 Oi} \,{}^R p_{3 Oi}\rangle |P_{e_H} m_H n_H\rangle\langle P_{e_H} m_H
n_H|\langle P_{e_{Oi}}, {}^L p_{3 Oi} \,{}^R p_{3 Oi}|,
\end{gather*}
the above density matrix is clearly diagonal.
If we then compute
\begin{gather*}
-\operatorname{Tr} (\rho^{\rm local}_{\rm reduced}\ln\rho^{\rm local}_{\rm reduced})  = \frac{ \beta
A_{\rm BH}}{4 l_p^2} + {\rm corrections},
\end{gather*}
where $\beta$ is a~prefactor~\cite{adg3}.
The exact computation is not obvious, but if we assume that every~$j_H$ is the same, then this
gives $-\operatorname{Tr}\big(\rho^{\rm local}_{\rm reduced}\ln\rho^{\rm local}_{\rm reduced}\big) =\sum_{v_{\rm
out}}\ln(2j_H+1)=A_H\ln(2j_H+1)/[l_p^2(j_H+1/2)]$ where we used the constraint~\eqref{eqn:area}
such that the $\sum(j_H+1/2)=A_H/l_p^2$.
For dif\/ferent distribution of spins one has to use combinatorics~\cite{adg2,adg3}.

Thus to leading order the entropy as computed from the density matrix is indeed Bekenstein--Hawking
modulo the $\beta$ which is a~function of the Immirzi parameter and other details of the
combinatorics of the graph can be set to~1.

\subsection{Quantum correlations}

\begin{figure}[t]
\centering
\includegraphics{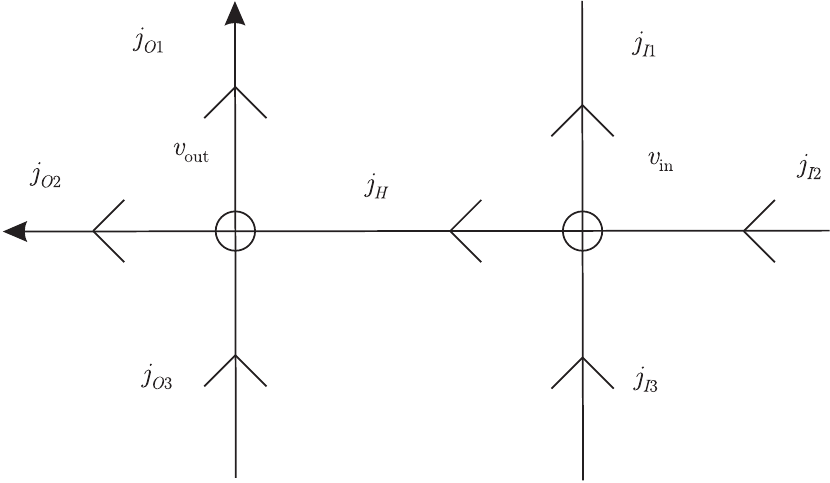}
\caption{Spins at the vertices.}
\label{fig:graph1}
\end{figure}

We now discuss the correlations across the horizon due to intertwiners placed at the vertices.
As evident the coherent states are def\/ined in a~${\rm SU}(2)$ Hilbert space.
For the holonomy variables, the ${\rm SU}(2)$ gauge transformations act at the beginning and end point of
the edges.
$h_e\rightarrow g^{-1}(v_1)h_e(v_1,v_2)g(v_2)$.
These changes will change the spin network state.
To make the state gauge invariant one imposes intertwiners at the vertices.
These are structures which map the spin networks at the vertex to the trivial representation.
Thus the state does not transform due to the intertwiners.
We concentrate on two vertices (suppress one dimension) and take each vertex to be four valent at
the horizon and observe the role of the intertwiners.
From the Fig.~\ref{fig:graph1}
two of the edges are ingoing, the other two
are outgoing at each vertex.
The gauge transformations
act on the holonomies and hence the basis states on each edge at the vertices.
Integrating (group averaging) over the gauge group at the vertices, one ensures gauge invariance~\cite{lqg,lqg1,lqg2}.
This solves the Gauss constraint.
We f\/ind that when done carefully this gives us correlations between the coherent state at inner and outer vertices
which are carried by the horizon edge which connects the two vertices.
This is therefore a~`quantum correlation'.
We describe some of the details
of the calculations here.
Let us take the two vertices as $v_{\rm in}$ and $v_{\rm out}$ and the spin labels $j_{I1}$, $j_{I2}$, $j_{I3}$, $j_H$ at the inner vertex and
$j_{O1}$, $j_{O2}$, $j_{O3}$, $j_H$ at the outer vertex.
Let us assume that at the inner vertex $j_{I1}$, $j_{H}$ are outgoing at that vertex, or
they `begin' at $v_{\rm in}$ and the $j_{I2}$, $j_{I3}$ are ingoing at the vertex or have their end
point at that vertex.
Similarly at the
outer vertex, $j_{H}$ and $j_{O3}$ are ingoing or end at $v_{\rm out}$ and $j_{O1}$, $j_{O2}$ are
outgoing and begin at the vertex $v_{\rm out}$.
To introduce the intertwiners we observe the basis spin network functions and their transformations.
The basis spin network states $|jmn\rangle $ at the inner vertex in the $h_e$ representation are given as
($\sqrt{2j+1}$ factor is not shown but it exists in each basis state for normalisation
of the inner product)
\begin{gather*}
\pi_{j_{I1}}(h)_{m_{I1}n_{I1}} \pi_{j_{I2}}(h)_{m_{I2}n_{I2}}\pi_{j_{I3}}(h)_{m_{I3} n_{I3}}
\pi_{j_H}(h)_{m_H n_H},
\end{gather*}
the gauge action $U(g)$ at the vertex is such that
\begin{gather*}
\pi_{j_{I1}} (gh)_{m_{I1} n_{I1}}\pi_{j_{I2}}\big(hg^{-1}\big)_{m_{I2}
n_{I2}}\pi_{j_{I3}}\big(hg^{-1}\big)_{m_{I3} n_{I3}}\pi_{j_H}(gh)_{m_H n_H}.
\end{gather*}
Upon group averaging one obtains a~projector onto the gauge invariant Hilbert space which written
using the $3jm$ symbols is thus (see Appendix~\ref{appendixA} for further details)
\begin{gather}
\sum_{\{j_k\} p_{I1} q_{I2} q_{I3} p_H}(2j_k+1)\left(
\begin{matrix} {j_{I 1}}
&j_{H}&j_k\\m_{I1}&m_H&m_k
\end{matrix}\right)
\left(
\begin{matrix} j_{I 1}&j_{H}&j_k\\p_{I1}&p_H&n_k
\end{matrix}\right)\nonumber \\
\qquad
{}\times
\overline{\left(
\begin{matrix} j_{I 2} &j_{I3}&j_k\\n_{I2}&n_{I3}&m_k
\end{matrix}\right)}
\overline{\left(
\begin{matrix} j_{I 2} &j_{I 3}&j_k\\q_{I2}&q_{I3}&n_k
\end{matrix}\right)} \nonumber \\
\qquad
{}\times
\pi_{j_{I1}}(h)_{p_{I1}n_{I1}} \pi_{j_{I2}}(h)_{m_{I2}q_{I2}}\pi_{j_{I3}}(h)_{m_{I3} q_{I3}}
\pi_{j_H}(h)_{p_H n_H}.\label{eqn:inner}
\end{gather}
We can then collect the basis vectors and the $3jm$ symbols contracted with them as a~$|\rm in\rangle $ basis
\begin{gather*}
= \sum_{\{j_k\}}
\left(
\begin{matrix}
{j_{I1}}&j_{H}&j_k\\
m_{I1}&m_H&m_k
\end{matrix}
\right)
\overline{
\left(
\begin{matrix}
{j_{I 2}} &j_{I3}&j_k\\
n_{I2}&n_{I3}&m_k
\end{matrix}\right)}
 |{\rm in}\rangle.
\end{gather*}

If we take $\psi_{j}(g_e)_{mn}=\frac1{\langle \psi|\psi\rangle}e^{-t/2j(j+1)}\pi_{j}(g_e)_{mn}$, the coherent
state wavefunction at the inner vertex is
\begin{gather*}
\sum_{\{j\}}\prod_{i=1}^{3}\psi_{j_{Ii}}(g_{e_{Ii}})_{m_{Ii} n_{Ii}}\left(
\begin{matrix} {j_{I
1}} &j_{H}&j_k\\m_{I1}&m_H&m_k\end{matrix}\right)\overline{\left(
\begin{matrix} {j_{I 2}}
&j_{I3}&j_k\\n_{I2}&n_{I3}&m_k
\end{matrix}\right)} |\rm in\rangle.
\end{gather*}

Note that this way, only the indices which begin or end at the $v_{\rm in}$ are gauge invariant,
the other end of the edges remains `free' and can be attached to
another vertex in an invariant way.
Exactly in the same way one can obtain the gauge invariant wavefunction at the outer vertex,
comprising of the $3jm$ symbols and the corresponding wavefunctions.
And
therefore one can def\/ine the intertwiners at the outer vertex but notice immediately the horizon
edge is shared by both the outer vertex and the innner vertex, and thus, the correct way to write
the coherent
state for the edges meeting at the two vertices is this
\begin{gather*}
|\psi(v_{\rm out},v_{\rm in})\rangle = \sum_{\{j\}}\Psi (\{j_{0i}\},\{j_H\},\{j_{Ii}\})|{\rm out}\rangle |
H \rangle |{\rm in}\rangle,
\end{gather*}
where $|H\rangle $ is the horizon state for the horizon edge crossing the horizon (see Fig.~\ref{fig:horizon}).
Clearly the coef\/f\/icient of the basis states do not factorise and can be identif\/ied as
\begin{gather*}
\Psi(\{j_{0i}\},\{j_H\},\{j_{Ii}\}) = \prod_{i=1}^3\psi_{j_{Oi}}(g_{e_{Oi}})_{m_{Oi} n_{Oi}}
\sum_{\tilde j_k'}\left(
\begin{matrix}
j_{0 1} &j_{O2}&{\tilde j_k'}\\m_{O1}&m_{O2}&{\tilde m_k'}
\end{matrix}
\right)
\\
\phantom{\Psi(\{j_{0i}\},\{j_H\},\{j_{Ii}\})=}
\times
\overline{\left(
\begin{matrix}
{\tilde j_{k'}}&j_{O3}&j_H\\
{\tilde m}_{k'}&n_{O3}&n_k
\end{matrix}
\right)}
\psi_{j_H}(g_{e_H})_{m_H n_H}
\\
\phantom{\Psi(\{j_{0i}\},\{j_H\},\{j_{Ii}\})=}
\times
\sum_{j_k}\prod_{i=1}^{3}\psi_{j_{Ii}}(g_{e_{Ii}})_{m_{Ii} n_{Ii}}\left(
\begin{matrix} {j_{I
1}} &j_{H}&j_k\\
m_{I1}&m_H&m_k
\end{matrix}\right)
\overline{\left(
\begin{matrix}
{j_{I 2}}&j_{I3}&j_k\\
n_{I2}&n_{I3}&m_k
\end{matrix}\right)}.
\end{gather*}
The density matrix is then
\begin{gather*}
\rho_{\rm inv}= |\psi(v_{\rm out},v_{\rm in})\rangle\langle \psi(v_{\rm out},v_{\rm in})|.
\end{gather*}

This in a~very sure way establishes that correlations exist across the horizon surface, the
question however is that is Bekenstein--Hawking entropy recovered from the tracing mechanism as
previously?
Ab initio it is not expected that this will happen, as these gauge invariant coherent states do not
have the appropriate peakedness properties of the gauge non-invariant
coherent states~\cite{thiemwinkl}.
We simply implement the tracing of the inner edges to f\/ind the reduced density matrix at this stage
and demonstrate the entanglement.

The reduced density matrix is easily obtained, using the trace over the inner basis states and some
$3jm$ symbol identities (see Appendix~\ref{appendixA}).
The matrix elements turn out to be
\begin{gather}
\sum_{\{j_{Ii}\}}
\Psi\left(\{j_{Oi}\},\{j_H\},\{j_{Ii}\})\bar\Psi(\{j'_{Ii}\},\{j'_H\},\{j_{Oi}\}\right)
\nonumber\\
\qquad{}
=\Psi(\{j_{0i}\},\{j_H\})\bar\Psi(\{j'_H\},\{j'_{Oi}\})\sum_{\{j_{Ii}\},j_k}\frac{(2
j_k+1)\delta(j_{I1}j_Hj_k)\delta(j_k j_{I2}j_{I3})}{(2j_H+1)} \nonumber
\\
\qquad\quad{}
\times
\prod_{i=1}^3\psi_{j_{Ii}}(g_{e_{Ii}})_{m_{Ii}
n_{Ii}}\prod_{i=1}^3\bar{\psi}_{j_{Ii}}(g_{e_{Ii}})_{m'_{Ii} n'_{Ii}}   \delta_{n'_{I1} n_{I1}}
\delta_{m'_{I2} m_{I2}}\delta_{m'_{I3} m_{I3}} \nonumber
\\
\qquad\quad{}
\times
\left(
\begin{matrix} {j_{I 1}}
&j_{H}&j_k\\m_{I1}&m_H&m_k
\end{matrix}\right)
\left(
\begin{matrix} {j_{k}} &j_{I2}&j_{I3}\\m_{k}&n_{I2}&n_{I3}
\end{matrix}\right)
\left(
\begin{matrix} {j_{I 1}} &j_{H}&j_k\\m'_{I1}&m'_H&m'_k
\end{matrix}\right)
\left(
\begin{matrix} {j_{k}} &j_{I2}&j_{I3}\\m'_{k}&n'_{I2}&n'_{I3}
\end{matrix}\right).\!\!\!\!\label{eqn:trace}
\end{gather}

Whereas it is not obvious what this will yield for the entropy which is $S_{\rm BH}=-\operatorname{Tr}(\rho\ln\rho)$, one can use the fact that again in the limit that $t\rightarrow0$,
the product of wavefunctions will assume the form in equation~\eqref{eqn:limit}
\begin{gather*}
\sum_{\{j_i\}}
\Psi\left(j_{Oi},j_H,j_{Ii})\bar\Psi(j_{Ii},j'_H,j'_{Oi}\right)
=\Psi(j_{0i},j_H)\bar\Psi(j'_H,j'_{Oi})
\sum_{j_k}\frac{(2 j_k+1)\delta(j_{I1}j_Hj_k)\delta(j_k j_{I2}j_{I3})}{(2j_H+1)}
\\
\qquad {}
\times
\left(
\begin{matrix} {j^{\rm cl}_{I 1}} &j_{H}&j_k\\m^{\rm cl}_{I1}&m_H&m_k
\end{matrix}\right)
\left(
\begin{matrix} {j_{k}} &j^{\rm cl}_{I2}&j^{\rm cl}_{I3}\\m_{k}&n^{\rm cl}_{I2}&n^{\rm
cl}_{I3}\end{matrix}\right)
\left(
\begin{matrix} {j^{\rm cl}_{I 1}} &j_{H}&j_k\\m^{\rm
cl}_{I1}&m'_H&m'_k
\end{matrix}\right)
\left(
\begin{matrix} {j_{k}} &j^{\rm cl}_{I2}&j^{\rm cl}_{I3}\\m'_{k}&n^{\rm cl}_{I2}&n^{\rm
cl}_{I3}
\end{matrix}\right).
\end{gather*}

The above reduced density matrix is precisely of the form expected to yield the entropy, but the
$3jm$ symbols are non-trivial and one can use some asymptotic form given in~\cite{asym} for large
$j$ to estimate the form
of the density matrix.
Details of this calculations and further computations of the entropy will appear
in~\cite{adg4}.
{\it This is the absolute proof of the fact that the coherent state introduced to describe the
black hole is entangled across the horizon and does not factorise into inside and outside
coefficients.}
There has been work in observing entanglement in semiclassical states independently as in~\cite{hustern}.

\section{Time evolution}
In this section we discuss time evolution of the system.
Fortunately as the system is semiclassical, a~natural
choice of time is the asymtotic time coordinate.
Using the Lemaitre slice and evolving the state in Lemaitre time would not serve the purpose as
that is
the frame of an infalling observer.
We use the frame of an observer stationed outside the horizon to evolve
the slice in her time as at the asymptotics this coincides with Minkowski time.
Again we concentrate on the evolution
of a~local patch surrounding the horizon.
Normally the vertices within the horizon are classically forbidden to
the outside observer.
However we allow for access to those vertices in the `quantum mechanical' evolution.
The
`Hamiltonian' which is used to evolve the horizon is the Brown and York quasi-local energy.

\subsection{Semi classical Hamiltonian}
Quasi-local Hamiltonian is def\/ined as the energy enclosed in a~f\/inite space.
This is obtained
using a~particular def\/inition~\cite{bro23}, as the `surface' integral of the extrinsic curvature
with which the
surface is embedded in three space.
This generates time translation in the timelines orthogonal to the spatial slice.
In
our case, we take the bounding surface to be the horizon and the quasilocal energy is given by
the surface term~\cite{bro23,haw23}
\begin{gather*}
\tilde H= \frac{1}{\kappa}\int d^2x \sqrt{\s} k,
\end{gather*}
where $k$ is the extrinsic curvature with which the 2-surface, which in this case is the horizon
$S^2$ embedded
in the spatial 3-slices, and $\s$ is the determinant of the two metric $\s_{\m\n}$ def\/ined on the
2-surface.
This `quasilocal energy' is measured with reference to a~background metric.
Thus $H=\tilde H-H_0$ where $H_0$
is the energy in the background.
We
concentrate on the physics observed by an observer stationed at a~$r= {\rm const}$ sphere.

The metric in static $r={\rm const}$ observer's frame is
\begin{gather*}
ds^2= - f^2 dt^2 + r^2 \big(d\th^2 + \sin^2\th d\phi^2\big),
\end{gather*}
 $f=\sqrt{1-r_g/r}$, where $r_g$ is the Schwarzschild radius.
If we take $n_{\m}$ to be the space-like vector,
normal to the 2-surface, then the extrinsic curvature is given by
$
k_{\m \n}= \s_{\m}^{\a}\nabla_{\a}n_{\n}
$
and the trace is obviously
\begin{gather}\label{eqn:trace1}
k= \nabla^{\a}n_{\a}.
\end{gather}
In the special slicing of the stationary observer the normal to the horizon 2-surface is given by
$(0,f(r),0,0)$.
However, we built the density matrix on the Lemaitre slice.
The Lemaitre and the Schwarzschild observer's
coordinates are related by the following coordinate transformations
\begin{gather*}
\sqrt{\frac{r}{r_g}} dr = (dR - d\tau), \qquad
dt = \frac{1}{1-f'}\left(d\tau - f' dR \right),\qquad f' = \frac{r_g}{r}.
\end{gather*}

The $r= {\rm const}$ cylinder of the Schwarzschild coordinate corresponds to $dR=d\tau$ of the Lemaitre
coordinates,
and for these $dt=d\tau$.
Thus unit translation in the $t$ coordinate coincides with unit translation in the
$\tau$ coordinate.
Further, the intersection of the $r=  {\rm const}$ cylinder with a~$t= {\rm const}$ surface
coincides with the intersection of $r={\rm const}$ and the $\tau={\rm const}$ surface.
Thus in the initial slice, the QLE Hamiltonian can be written
in terms of the Christof\/fel symbols of the Lemaitre slice.
Even though the
generic transformation of the Christof\/fel symbols from one coordinate frame to the other is
inhomogeneous,
in this example, the transformation is trivial.
The trace of the extrinsic curvature~\eqref{eqn:trace1} then written using Christof\/fel symbols
turns out to~be
\begin{gather}\label{hamil1}
H= \frac1{2 \kappa}\int d\theta d\phi \sqrt{g_{\th \th} g_{\phi \phi}} \left[-g^{\th \th}\frac{\partial
g_{\th \th}}{\partial r} - g^{\phi \phi} \frac{\partial g_{\phi \phi}}{\partial r}\right]f(r) - H_0.
\end{gather}
The reference frames' quasilocal energy $H_0$ is a~number, it just def\/ines the zero point
Hamiltonian.
Thus, we replace the classical expressions by operators evaluated at the $\tau= {\rm const}$ slice.
In the f\/irst approximation
we simply take the $f(r)$ as classical $\sqrt{1-r_g/r}=\sqrt{\delta_{e_r}/2r_g}=\epsilon$, as this
arises due to the coordinate transformation and the norm of the vector $n_r$ in the previous frame.
In the re-writing of~\eqref{hamil1} in regularised LQG variables the Hamiltonian
appears rather complicated.

One can rewrite these in a~much simpler form,
using the apparent horizon equation.
Since the Hamiltonian is an integral over the horizon, the variables will satisfy the apparent
horizon equation~\eqref{classical} up to quantum f\/luctuations.
Thus the Hamiltonian operator is then re-written~as
\begin{gather*}
H_{\rm horizon}= \frac \e{\kappa}\int d\theta d\phi \sqrt{g_{\th \th} g_{\phi \phi}}
\big[K^I_{\theta}e^{I\theta} + K^I_{\phi} e^{I \phi}\big],
\end{gather*}
where we have used the classical apparent horizon equation~\eqref{classical} (with $q^{rr}=1$).
The regularised LQG version of this is
\begin{gather}\label{hamil2}
H_{\rm horizon} = \frac{C a~\e}{2\kappa \delta_{e_\th}s_{e_\th}} \sum_{v_{\rm out}}\operatorname{Tr}\left[h_{e_\th}^{-1} T^I \beta\frac{\partial}{\partial \beta} h_{e_\theta}\right] P_{e_\theta}^I + \text{h.c.}
+ (\th\rightarrow\phi),
\end{gather}
where $C$ consists of some dimensionless constants $s_{e_\th}$ is the 2-dimensional area bit over
which~$E^{\theta}_I$ is smeared, $a$ is a~dimensionfull constant which appears to get the $P_{
e_\theta}^I$ dimension less.
$\delta_{e_{\th}}$ is the length for the
angular edge $e_{\th}$ over which the gauge connection is integrated to obtain the holonomy.
The sum over $v_{\rm out}$ is the set of vertices immediately
outside the horizon.
The~\eqref{hamil2} can then be lifted to an operator.
However, this Hermitian operator gives no change to horizon entropy as the coherent state is
evolved from one time slice to the next.
Surprisingly a~non-Hermitian term evolves the slices non-trivially.
These anti-Hermitian terms
can be easily found in this formalism if we allow for the vertices within the horizon to be exposed
to the outside observer.
In this case as the region is within the classical horizon, the norm of the Killing vector
is negative, and $n_r$ has components which are imaginary.
The $\epsilon\rightarrow\pm\iota\epsilon$.
One sees why the inner vertices are relevant in a~semiclassical evolution even if they are
classically forbidden: In quantum mechanics tunneling is a~well known phenomena, and inclusion of
the inner vertices facilitates the process.

Thus the contribution of `inner vertices' $v_{\rm in}$ is added to the Hamiltonian and it gives the
anti-Hermitian contribution to the Hamiltonian.
Thus $H_{\rm horizon}=\frac12\big[\sum_{v_{\rm out}}H_{v_{\rm out}}+\sum_{v_{\rm in}}H_{v_{\rm
in}}\big]$.
The regularised Hamiltonian is not Hermitian, and the evolution equation is
\begin{gather}\label{ch}
\iota \hbar \frac{\partial \rho}{\partial \tau} = H \rho - \rho H^{\dag}.
\end{gather}

Thus one takes the evolved slice to have a~density matrix
\begin{gather*}
\rho= \rho^0 -\frac{\iota \delta \tau}{\hbar}\big[H\rho^0-\rho^0H^{\dag}\big].
\end{gather*}

The entropy computed using this density matrix is dif\/ferent from the entropy in the previous slice
and the dif\/ference of entropy can be seen as obtained using a~simplif\/ied model.

Let the density matrix be def\/ined for a~system whose states are given in the tensor product Hilbert
spaces $H_1\otimes H_2$
and given by
\begin{gather*}
|\psi\rangle= \sum_{ij} d_{i j} |i\rangle |j\rangle,
\end{gather*}
where $|i\rangle $ is the basis in $H_1$ and $|j\rangle $ is the basis in $H_2$ and $d_{ij}$ are the
non-factorisable coef\/f\/icients of the wavefunction
in this basis.
Let us label the wavefunction at time $t=0$ to be given by the coef\/f\/icients $d^{0}_{ij}$.
The density matrix is
\begin{gather*}
\rho^{0}= \sum_{ ij i' j'} d^{0 *}_{i' j'} d^{0}_{i j} |i\rangle |j\rangle\langle j'|\langle i'|.
\end{gather*}
The reduced density matrix if one traces over $H_2$ is
\begin{gather*}
\operatorname{Tr}_2\rho^{0}= \sum_{i i'}\sum_j d^{0 *}_{i' j} d^0_{i j}|i\rangle\langle i'|.
\end{gather*}
We now evolve the system using a~Hamiltonian which has the matrix elements $H_{i j i'
j'}|i\rangle |j\rangle\langle j'|\langle i'|$, we
assume that the Hamiltonian does not factorise, that is there exists interaction terms between the
two Hilbert spaces.
The evolution equation is
\begin{gather*}
i \hbar \frac{\partial \rho}{\partial \t}= [ H, \rho],
\end{gather*}
which in this particular basis gives the density matrix elements at an~inf\/initesimally nearby slice to be
\begin{gather*}
d^{\delta\t *}_{i' j'} d^{\delta \t}_{i j} = d^{0 *}_{i' j'} d^{0}_{i j}
-\frac{i}{\hbar} \delta \t \left[\sum_{k l} \left(H_{ij kl} d^{0}_{k l} d^{0 *}_{i' j'} - d^{0}_{i
j} d^{0 *}_{kl} H_{kl i' j'}\right)\right].
\end{gather*}
Thus we evolve the `unreduced' density matrix and then trace over the $H_2$ in the evolved slice.
The reduced density matrix in the evolved slice is
\begin{gather*}
\sum_j d^{\delta\t *}_{i' j} d^{\delta \t}_{i j} = \sum_j d^{0 *}_{i' j} d^{0}_{i j} -
\frac{i}{\hbar} \delta \t \left[\sum_{k l j } \left(H_{ij kl} d^{0}_{k l} d^{0 *}_{i' j} - d^{0}_{i
j} d^{0 *}_{kl} H_{kl i' j}\right)\right].
\end{gather*}
This gives
\begin{gather*}
\rho^{\delta \t} = \rho^{0} -\frac{i}{\hbar} \delta \t A,
\end{gather*}
where $A$ represents the commutator.
Clearly the entropy in the evolved slice evaluated as $S_{\rm BH}^{\delta\t}=-\operatorname{Tr}(\rho\ln\rho)$
can be found as
\begin{gather*}
S_{\rm BH}^{\delta \t} = S_{\rm BH}^{0} +\frac{i}{\hbar} \delta \t
\big[\operatorname{Tr} A \ln \rho^0 + \operatorname{Tr}\rho^{0}\rho^{0 \ -1} A\big].
\end{gather*}

Given the def\/inition of $A_{ii}$, one gets
\begin{gather*}
A_{ii} = \sum_{jkl} \left[\rho^0_{ijkl}H_{klij}- H_{ijkl}\rho^0_{klij}\right].
\end{gather*}
In case both the Hamiltonian and the density operator are Hermitian, one obtains
\begin{gather*}
\sum_j A_{jj}= 2 \iota   \operatorname{Im} \operatorname{Tr}\big(\rho^0 H\big).
\end{gather*}

This is clearly calculable, and gives the change in entropy $\Delta S_{\rm BH}$.
The $\ln\rho^0$ term yields corrections, and we ignore it in the f\/irst approximation.
However if the Hamiltonian is non-Hermitian one gets
\begin{gather}\label{eqn:change}
\Delta S_{\rm BH} = \frac{\iota \delta \tau}{\hbar} {\rm Tr}\big[H \rho^0 - \rho^0 H^{\dag}\big]
\end{gather}
instead of the commutator.

In case of the gravitational Hamiltonian, the above can be computed using a~${\rm U}(1)$ projection thus.
Let us take the ${\rm U}(1)$ case to make the calculations easier and observe the action of the Hamiltonian
on the evolution of the coherent state.
The spin network states are replaced by $|n\rangle =e^{\iota n\zeta}$, $0<\zeta<2\pi$, $n$ is an integer
and the coherent states are
\begin{gather*}
\psi^t(g_e)= \sum_n e^{-(t n^2)/2} e^{i n(\chi_e - ip_e)} e^{-\iota n \zeta}.
\end{gather*}
$g_{ne}=e^{i n(\chi_e-i p_e)}$ is the complexif\/ied phase space element in the `$n$th' representation.

The quasi local energy operator~\eqref{hamil2} also takes the simplif\/ied form
\begin{gather}\label{hamil}
H_{\rm horizon}^{\rm U(1)} = -\frac12 C'\iota \hat h_{e}^{-1}\beta \frac{\partial}{\partial \beta}
\hat h_e \hat p_e
+ \frac12 C'\iota \hat p_e \beta \frac{\partial \hat
h_e^{-1}}{\partial \beta} \hat h_e.
\end{gather}

The prefactors have been clubbed into $C'$.
The U(1) coherent states are eigenstates of an annihilation operator def\/ined thus:
\begin{gather*}
\hat g_e = e^{t/2} e^{\hat p_e}  \hat h_e, \qquad \hat g_e |\psi\rangle = g_e |\psi\rangle.
\end{gather*}

The holonomy operator can thus be written as
\begin{gather*}
\hat h_e = e^{-t/2} e^{-\hat p_e}  \hat g_e.
\end{gather*}
And the derivative w.r.t.\ Immirzi parameter of the holonomy which appears in the def\/inition of the
Hamiltonian replaced by
\begin{gather*}
\beta\frac{\partial \hat h_e}{\partial \beta} =  e^{-t/2} \left[ -\beta \frac{\partial \hat
p}{\partial \beta} e^{- \hat p_e}   \hat g_e + e^{-\hat p_e}   \beta\frac{\partial \hat
g_e}{\partial \beta}\right]
= e^{-t/2} \left[ \hat p_e   e^{-p_e}   \hat g_e + e^{-\hat p}  \beta\frac{\partial \hat g_e
}{\partial \beta} \right].
\end{gather*}

The dependence of the operator $p$ on the Immirzi parameter is known~\eqref{defn}, and thus we
could evaluate the
derivative ($\beta\partial_{\beta}p_e(\beta)=\beta\partial_{\beta}(p_e(1)/\beta)=-p_e(\beta)$).

The term
\begin{gather}\label{tr}
\operatorname{Tr}\big(\rho^0 H^{\rm U(1)}_{\rm horizon}\big)
\end{gather}
is then computable.
Let us take the f\/irst term of~\eqref{hamil} and f\/ind \eqref{tr}.
As $\rho^{0}=|\psi\rangle\langle \psi|$,~\eqref{tr} gives simply (we drop the `$e$' label for brevity)
\begin{gather*}
\langle \psi|H^{\rm U(1)}_{\rm horizon} |\psi\rangle =
\langle \psi|-\frac12 C'\iota \hat h^{-1}\beta \frac{\partial}{\partial \beta} \hat h   \hat p|\psi\rangle
+ \langle \psi|\text{h.c.}|\psi\rangle
\\
\phantom{\langle \psi|H^{\rm U(1)}_{\rm horizon} |\psi\rangle}
=-\frac12 \iota C' \langle \psi| \hat g^{\dag} e^{-t/2} e^{-\hat p} e^{-t/2}\left[\hat p\ e^{-\hat p} \hat g
+ e^{-\hat p} \b \frac{\partial \hat g}{\partial \beta} \right]\hat p|\psi\rangle + \langle \psi|\text{h.c.}|\psi\rangle
\\
\phantom{\langle \psi|H^{\rm U(1)}_{\rm horizon} |\psi\rangle}
=-\frac12 \iota C' e^{-t} g^* \left[\langle \psi|e^{-\hat p}  \hat p   e^{-\hat p}\hat p   |\psi\rangle g
+\langle \psi| e^{-2 \hat p}  \b\frac{\partial \hat g}{\partial \beta}\hat p   |\psi\rangle\right]
+\langle \psi| \text{h.c.} |\psi\rangle.
\end{gather*}

We then concentrate on the 2nd term of the above
\begin{gather*}
\langle \psi|e^{- 2\hat p}\b\frac{\partial \hat g}{\partial \beta}\hat p |\psi\rangle
= \langle \psi|e^{-2 \hat p}\b\frac{\partial \hat g}{\partial \beta}
  \int\! d\nu(g') |\psi'\rangle\langle \psi'| \hat p
|\psi\rangle
= \!\int\! d \nu (g') \b \frac{\partial g'}{\partial \beta} \langle \psi| e^{-2 \hat p}|\psi'\rangle\langle \psi'|\hat p
|\psi\rangle,
\end{gather*}
where we have used the fact that coherent states resolve unity.
It can be shown that the expectation value of the
operators in the $t\rightarrow0$ collapses the integral to $g'=g$ point~\cite{thiemwinkl}.
Thus one obtains from the above
\begin{gather}
\operatorname{Tr}\big(\rho^0 H^{\rm U(1)}_{\rm horizon}\big) =  - \frac12 \iota C' e^{-t/2} \left[ p + g^* e^{-2 p}
\b\frac{\partial g}{\partial \beta}\right]p + \text{h.c.}
= C' \b \frac{\partial \chi}{\partial \beta} p, \label{expect}
\end{gather}
which is real,  and thus
\begin{gather*}
\Delta S_{\rm BH}=0
\end{gather*}
(this is actually the classical quasi local energy as it should be from $\operatorname{Tr}(\rho^0H_{\rm
horizon})$).

This is obvious, as the way the Hamiltonian is def\/ined, this is simply a~function of the Hilbert
space
outside the horizon, and the matrix elements of this will not yield anything new.
We approximated the horizon sphere by summing over
$v_1$ vertices immediately outside the horizon.
We could do the same by summing over $v_2$
vertices immediately within the horizon.
For the Lemaitre slice, the metric is smooth at the
horizon, and one can take the `quantum operators' evaluated at the vertex~$v_2$.
In this case however, as the region is within the classical horizon, the norm of the Killing vector
is negative, and~$n_r$ has components which are imaginary.
The $\epsilon\rightarrow\pm\iota\epsilon$.
Thus $H_{\rm horizon}=\frac12[\sum_{v_1}H_{v_1}+\sum_{v_2}H_{v_2}]$.
In the evaluation of the quasi local energy term, the energy would emerge correct in the $
\d_{e_r}\rightarrow0$ limit as $\e\rightarrow0$.
The regularised Hamiltonian is not Hermitian, and the evolution equation is~\eqref{ch}.

The same type of expectation or trace is obtained of the imaginary term (or $v_2$) is obtained as
in~\eqref{expect}, but this does not cancel in \eqref{eqn:change}, and
\begin{gather*}
\Delta S_{\rm BH}= \mp \frac{\delta \tau}{ \hbar} C' \beta\frac{\partial \chi}{\partial \beta} \ p.
\end{gather*}
The `rate of change' of entropy is thus
\begin{gather*}
\dot \Delta S_{\rm BH}= \mp \frac{\tilde C}{l_p^2} \beta \frac{\partial \chi}{\partial \beta} \ p
\end{gather*}
(we extracted the $\kappa$ from $C'$ to get $l_p^2$ and rewrote the rest of the constants as~$\tilde C$).

Thus there is a~net change in entropy,
but, to see if this is Hawking radiation, we have to couple matter to the system.

In the case of the black hole system, due to spherical symmetry, in the classical limit, from the
regularisation of the extrinsic curvature components $K_{\theta,\phi}^I$ which appear in the Brown
and York quasi-local energy and its LQG regularisation~\eqref{hamil2}, the change of entropy is
\begin{gather*}
\Delta S_{\rm BH}= \mp \frac{\tilde C\epsilon \delta \tau}{l_p^2} \sum_{v}
\beta\left[\frac{\partial \chi_{e_\th}}{\partial \beta} P_{e_\th} + \frac{\partial
\chi_{e_\phi}}{\partial \beta}P_{e_\phi}\right].
\end{gather*}
Due to the nature of the classical metric, the $P_{e_{\th,\phi}}^I$ can be gauge f\/ixed to have one
non-zero component in the internal directions~\cite{adg5}.
Let that be some f\/ixed index $r$, then $P_{e_{\th,\phi}}=\delta^r_I P_{e_{\th,\phi}}^I$.
In the computation of the Hamiltonian then it can be shown that if the holonomy is assumed to be
$h_e=e^{i\chi^I T^I}$ then, the $r$ component of $\chi^r$ contributes to the Hamiltonian, and
this is what we have set as $\chi_{e_{\th,\phi}}$ in the above.
From~\cite{adgcoh} one f\/inds the $\chi_{e_\th}$ and the derivative of that at the horizon is a
function of $\sqrt{r_g/r}$ which gives a~constant.
The $P_{e_\th}$ are the regularised densitised triads at the horizon and thus they give area of the
two surfaces comprising the decomposition of the three sphere in the angular dimensions.
As it is evident,
\begin{gather*}
P_{e_\th}=\frac{1}{a}\int * E_{\th}= \frac1a\int* r\sin\theta.
\end{gather*}
Thus they will be proportional to $r_g\sin\theta$ at the horizon.
Similar analysis is done for the $\phi$ sector
\begin{gather*}
\Delta S_{\rm BH}= \pm \frac{2\tilde C'\epsilon \delta \tau}{l_p^2} \sum_{v} dA_v \beta r_g,
\end{gather*}
where $dA_v$ is the area of the element of the two surface at the vertex $\sin\theta$ and $\tilde
C'$ has been redef\/ined to accomodate the regularisation constants which appear in $P_{e_\th}$.

Thus using the fact that $T_H=1/4\pi r_g$ one could def\/ine the `rate of entropy change' $\Delta
S_{\rm BH}/\delta\tau$ as
\begin{gather*}
\Delta \dot S_{\rm BH} \propto \pm \frac{\epsilon}{T_H}.
\end{gather*}
Note that this is change in entropy due to emission of one Hawking quanta of a~particular
frequency.
It should not be confused with entire black hole decay rate.

\subsection{Hawking radiation}
The fact that entropy change occurs due to Hamiltonian evolution which includes a~non-Hermitian
term is very interesting.
Hawking radiation however is a~f\/lux of particles emerging
from the horizon, and thus we have to trace the origin of such a~f\/lux.
In the previous section we found that as the system evolves in time, the horizon f\/luctuates and the
area decreases.
But is this Hawking radiation? Adding matter to a~`coherent state' description
of semiclassical gravity has been discussed~\cite{thiemsahl}.
Thus, given a~massless scalar f\/ield Lagrangian coupled to gravity, whose Hamiltonian
is given by
\begin{gather*}
H_{\rm sc}= \int d^3x \left[\frac{ \pi^2}{\sqrt{q}} + (\nabla \phi)^2 \right],
\end{gather*}
the `gravity' in the Hamiltonian can be regularised in terms of the $h_e$, $P^I_e$ operators in the
coherent state formalism.
The integral over the three volume
gets converted to a~sum over the vertices dotting the region.
Thus
\begin{gather*}
H_{\rm sc}^{v} = \sum_{v} H_{v} \big(h_e, P^I_e, V\big).
\end{gather*}

This Hamiltonian is an operator, and one evaluates an expectation value of the Hamiltonian in the
density matrix by computing the
trace of the product of the density matrix and the Hamiltonian
\begin{gather*}
\operatorname{Tr}\big(\rho^{\tau} H^{\tau}_{\rm sc}\big).
\end{gather*}

This Hamiltonian and the density matrix are then both evolved according to the time-like observers frame.
One gets
\begin{gather*}
i \hbar \frac{\partial H_{\rm sc}}{\partial \tau}= [H, H_{\rm sc}].
\end{gather*}
Thus one can compute the Hamiltonian in a~slice inf\/initesimally close to the previous slice $\tau
+\delta\tau$.
Using that,
\begin{gather*}
\operatorname{Tr}\big(\rho^{\tau+ \delta \tau} H^{\tau + \delta \tau}_{\rm sc}\big) - \operatorname{Tr}
\big(\rho^{\t} H_{\rm sc}^{\t}\big)= -(\d \t)^2 \operatorname{Tr} \{[H,\r^{\t}] [H, H_{\rm sc}^{\t}]\}.
\end{gather*}

It is very clear thus that the order $\d\t$ terms are zero for this.
However, allowing for the non-unitary evolution using the non-Hermitian
Hamiltonian, the $\d\t$ terms survive.
In fact the terms are
\begin{gather}
\operatorname{Tr}\big(\rho^{\tau+ \delta \tau} H^{\tau + \delta \tau}_{\rm sc}\big) - \operatorname{Tr}
\big(\rho^{\t} H_{\rm sc}^{\t}\big)\nonumber\\
\qquad{}
= -\frac{\iota \d \t}{\hbar} \operatorname{Tr}\big[(H \rho - \rho H^{\dag})H_{\rm sc}\big]
 - \frac{\iota \d \t}{\hbar} \operatorname{Tr}\big[\rho \big(H H_{\rm sc} - H_{\rm sc} H^{\dag}\big)\big].
\label{eqn:evol}
\end{gather}
In the exact Hilbert space of the matter gravity system, the states are not exactly in the tensor
product form, however, in the f\/irst approximation we take the states of the matter gravity system
to be $\prod_v|\psi(g)\rangle \otimes|\omega\rangle $ where $|\omega\rangle $ is a~matter state with particles of
energy~$\omega$.
At this step for simplicity we assume that the outer and inner vertex scalar state is at the same
energy $\omega$.
It is obvious that the f\/irst term in~\eqref{eqn:evol} gives the change in entropy as per~\eqref{eqn:change} times the matter Hamiltonian's eigenvalue $\Delta S_{\rm BH}\omega$.
The `rate' of particle creation thus has the form
\begin{gather*}
-\frac{\epsilon   \omega}{T_H},
\end{gather*}
where $T_H$ is the Hawking temperature for the signs $+(-)\iota\epsilon$ and negative (positive)
$\omega$.
The $\epsilon$ can be interpreted as the redshift factor.
If a~n particle matter state is taken, then the number will change accordingly.
Thus a~f\/lux of particles indeed emerges from the horizon.
To obtain a~complete f\/inite f\/lux one has to compute the f\/lux production over a~f\/inite
number of such time steps.
One can see that eventually a~exponential will emerge which is the required Boltzmann factor.
Further details can be found in~\cite{adg5}.

\subsection{Irreversible systems}

It has been known from the time of Boltzmann that entropic systems usually have irreversible time
evolution.
The Boltzmann H-theorem uses a
particular collision term in the evolution equation whose behavior gives rise to this irreversible
f\/low.
In complex systems physicists
have been using dif\/ferent techniques to create irreversibility in the dynamics.
In~\cite{prigogine}
this f\/low of a~complex system is formulated as
\begin{gather*}
\iota \frac{\partial \rho}{\partial t}= \big(\Phi^0 +\Phi^e\big)\rho,
\end{gather*}
where the usual Liouville operator has been written as broken into two pieces, one $\Phi^0$ which
is the reversible part and
the other $\Phi^e$ which is irreversible and non-Hermitian.
The $\Phi^e$ creates the entropy production during the f\/low of the system.
Thus it is remarkable that in the derivation of the time f\/low in presence of the horizon,
precisely such a~splitting occurs in the time-evolution equation.
It seems quantum gravity
should be formulated such that its dynamics should not be restricted to unitary operators.
A complete re-formulation of quantum gravity using the language of complex systems and
verif\/ication of some of the existing results is work in progress.
This will facilitate the description of the system to describe black hole evaporation.
Notice that as
the non-Hermitian component of the Hamiltonian
contributes from within the horizon, if all the mass of the black hole is radiated away, the term
within
the horizon will
also disappear, and the evolution will become unitary again.
However this as of yet speculative, a~stable
Planck size remnant might also form in the process.

\section{Conclusion}

Thus in this article we discussed coherent states for non-rotating black holes.
We then showed that these states can be used to explain the origin of entropy due to classical
correlations and then gave a~brief introduction
to a~computation of `quantum correlations' using ${\rm SU}(2)$ intertwiners at the vertices.
I then derived a~time evolution with a~non-Hermitian
Hamiltonian which generated entropy and this seems to create a~f\/lux of matter which escapes the
horizon.
This semiclassical derivation is
similar to the process of entropy production observed in complex systems.
However, the non-unitary nature of the evolution equation
might be due to the use of semiclassical `time' and an evolution operator which generates evolution in that time.
One has to further work with a~`quantum Hamiltonian' to verify these results.
Of course this brings us to the
problem of identifying time in quantum gravity, and the evolution in physical time remains a
project of research.
There is promise from the reduced `spherically symmetric' solution for the true Hamiltonian which
can be found in the following papers~\cite{sph,sph1,sph12}.
Other examples of reduced phase space spherically symmetric quantisation and studies of Hawking
radiation can be found in~\cite{sph2,sph21}.
For
the time being, reformulating quantum gravity in an attempt to answer all the questions is work in progress.

\appendix
\section{Intertwiners}\label{appendixA}

To build the `gauge invariant coherent states' we use the techniques of group averaging, the
details of which can be found in reviews of loop quantum gravity~\cite{lqg,lqg1,lqg2}.
For this the graph and its vertices are specif\/ied and the intertwiners at each vertex introduce the
gauge invariance.
For us, it is suf\/f\/icient to take a~graph which is comprised of vertices immediately outside the
horizon and those immediately inside the horizon, connected by a~radial edge.
If we suppress one angular dimension of the spherical horizon, and open up this graph, it will
appear as a~`ladder' structure.

If we isolate one rung of the graph and concentrate on that one obtains these two verti\-ces~$v_{\rm
out}$ connected to $v_{\rm in}$ with four edges meeting at
each vertex (see Fig.~\ref{fig:graph1}).
From the f\/igure it is clear that two of the edges are ingoing and two are outgoing at each vertex.
To obtain the
intertwiners at each of the vertex, we write the basis states.
At vertex $v_{\rm in}$ the basis states are (suppressing the $\sqrt{2j+1}$ factor which makes the
inner product~1)
\begin{gather*}\pi_{j_{I1}}(h)_{m_{I1}n_{I1}}\pi_{j_{I2}}(h)_{m_{I2}n_{I2}}\pi_{j_{I3}}(h)_{m_{I3}
n_{I3}}\pi_{j_H}(h)_{m_{H} n_{H}}
\end{gather*}
Implementing a~Gauge transformation at $v_{\rm in}$ one obtains
\begin{gather*}
\pi_{j_{I1}}(g h)_{m_{I1}n_{I1}}\pi_{j_{I2}}\big(hg^{-1}\big)_{m_{I2}
n_{I2}}\pi_{j_{I3}}\big(hg^{-1}\big)_{m_{I3} n_{I3}}\pi_{j_H}(g h)_{m_{H} n_{H}}
\\
\qquad{}
=\pi_{j_{I1}}(g)_{m_{I1}p_{I1}}\pi_{j_{I1}}(h)_{p_{I1}n_{I1}}\pi_{j_{I2}}(h)_{m_{I2}
q_{I2}}\pi_{j_{I2}}\big(g^{-1}\big)_{q_{I2}n_{I2}}
\\
\qquad\quad{}
\times
\pi_{j_{I3}}(h)_{m_{I3}q_{I3}}
\pi_{j_{I3}}\big(g^{-1}\big)_{q_{I3} n_{I3}}
\pi_{j_H}(g)_{m_{H} p_{H}}\pi_{j_H}(h)_{p_H n_H}.
\end{gather*}
Integrating over the group element using the measure $\int{\cal D}(g)$ then gives the $3jm$ symbols.
This then def\/ines the projector onto the gauge invariant space.
If we isolate the integrals this is
\begin{gather*}
\int {\cal D} g \ \pi_{j_{I1}}(g)_{m_{I1} p_{I1}}\pi_{j_{I2}}\big(g^{-1}\big)_{q_{I2}
n_{I2}}\pi_{j_{I3}}\big(g^{-1}\big)_{q_{I3} n_{I3}}\pi_{j_H}(g)_{m_H p_H}.
\end{gather*}
Inserting a~delta function using Peter--Weyl theorem
\begin{gather*}
\sum_{j_k m_k n_k} \ (2j_k+1)\pi_{j_k}(h)_{m_k n_k}\pi_{j_k}\big(h'^{-1}\big)_{m_k n_k} = \delta(hh'^{-1}).
\end{gather*}
One obtains
\begin{gather}
\sum_{j_k m_k n_k}\ (2 j_k+1)\int{\cal D}g\ \pi_{j_{I1}}(g)_{m_{I1}p_{I1}}\pi_{j_H}(g)_{m_H p_H} \pi_{j_k}(g)_{m_k n_k}
\nonumber \\
\qquad{}
\times
\int {\cal D} g' \pi_{j_k}\big(g'^{-1}\big)_{m_k
n_k}\pi_{j_{I2}}\big(g'^{-1}\big)_{q_{I2} n_{I2}}\pi_{j_{I3}}(g'^{-1})_{q_{I3} n_{I3}},
\label{integ}
\end{gather}
these integrate to $3jm$ coef\/f\/icients as per the
def\/inition~\cite{var}
\begin{gather*}\int{\cal D}g\pi_{j_1}(g)_{m_1n_1}\pi_{j_2}(g)_{m_2n_2}\pi_{j_3}(g)_{m_3n_3}=
\left(
\begin{matrix}j_1&j_2&j_3\\m_1&m_2&m_3
\end{matrix}\right)\left(\begin{matrix}j_1&j_2&j_3\\n_1&n_2&n_3
\end{matrix}\right).
\end{gather*}
Writing $\pi_{j}\big(g^{-1}\big)_{mn}=\pi_j(g^*)_{nm}=\pi^*_j(g)_{nm}$ in the second integral of~\eqref{integ}, one gets the complex conjugate of the $3jm$ symbols def\/ined above.
Thus the contribution to the inner vertex is as mentioned in~\eqref{eqn:inner}.

\section{Tracing}
We brief\/ly describe the tracing mechanism at the Inner vertex.
To trace over the basis states of $|j_{Ii}m_{Ii}n_{Ii}\rangle \equiv
\sqrt{2j_{Ii}+1}\pi_{j_{Ii}}(h)_{m_{Ii}n_{Ii}}$
we simply use the inner product of $\langle jmn|j'm'n'\rangle =\delta_{j j'}\delta_{m m'}\delta_{n n'}$.
Once the Kronecker delta functions have been implemented; the wavefunction at the inner vertex is
of the form
\begin{gather*}
\sum_{\{j\}}(2j_k + 1)\left(
\begin{matrix}j_{I1}&j_H&j_k\\m_{I1}&m_{H}&m_{k}
\end{matrix} \right)\left(
\begin{matrix}j_{I1}&j_H&j_k\\p_{I1}&p_{H}&n_{k}
\end{matrix} \right)
\overline{\left(
\begin{matrix}j_{k}&j_{I2}&j_{I3}\\m_{k}&n_{I2}&n_{I3}
\end{matrix} \right)}\overline{\left(
\begin{matrix}j_{k}&j_{I2}&j_{I3}\\n_{k}&q_{I2}&q_{I3}
\end{matrix} \right)}
\nonumber \\
\qquad{}
\times
\prod_{i=1}^3 \psi_{j_{Ii}}(g)_{m_{Ii} n_{Ii}}\prod_{i=1}^3 \bar{\psi}_{j_{Ii}}(g)_{m'_{Ii}
n'_{Ii}}\delta_{n'_{I1} n_{I1}}\delta_{m'_{I2}m_{I2}}\delta_{m_{I3} m'_{I3}}
\psi_{j_H}(g_H)_{m_H n_H}\bar\psi_{j'_H}(g_H)_{m'_H n'_H}
\nonumber \\
\qquad{}
\times
\sum_{j_k'}(2j_k'+1)\overline{\left(
\begin{matrix}j_{I1}&j'_H&j_k\\m'_{I1}&m'_{H}&m_{k'}
\end{matrix}\right)}\overline{\left(
\begin{matrix}j_{I1}&j'_H&j_{k'}\\p_{I1}&p'_{H}&n_{k'}
\end{matrix}\right)}
\nonumber \\
\qquad{}
\times
\left(
\begin{matrix}j_{k'}&j_{I2}&j_{I3}\\m_{k'}&n'_{I2}&n'_{I3}
\end{matrix}\right)\left(\begin{matrix}j_{k'}&j_{I2}&j_{I3}\\n_{k'}&q_{I2}&q_{I3}
\end{matrix}\right).
\end{gather*}
We then sum the $3jm$ symbols with common entries, e.g.
\begin{gather*}\sum_{q_{I2}
q_{I3}}\overline{\left(
\begin{matrix}j_{k}&j_{I2}&j_{I3}\\n_{k}&q_{I2}&q_{I3}
\end{matrix}\right)}
\left(
\begin{matrix}j_{k'}&j_{I2}&j_{I3}\\n_{k'}&q_{I2}&q_{I3}
\end{matrix}\right)=\frac{\delta_{j_k j'_k} \delta_{n_k n'_k}}{(2j_k+1)}\{j_k j_{I2} j_{I3}\},
\end{gather*}
where $\{j_1j_{2}j_{3}\}$ is the $3j$ symbol and is such that it is 1 if $j_1+j_2+j_3$ is an
integer and $|j_1-j_2|\leq j_3\leq j_1+j_2$, and 0 otherwise.
This process then f\/inds equation~\eqref{eqn:trace}.

\subsection*{Acknowledgements}
This work was supported by NSERC and University of Lethbridge Research Fund.

\pdfbookmark[1]{References}{ref}
\LastPageEnding

\end{document}